\documentclass[journal,comsoc]{IEEEtran}
\usepackage[T1]{fontenc}
\usepackage{algorithmic}
\usepackage[titlenumbered,ruled]{algorithm2e}
\usepackage{caption}
\usepackage[flushleft]{threeparttable}
\usepackage{cite}

\ifCLASSINFOpdf
\else
\fi
\usepackage{amsmath}
\usepackage[cmintegrals]{newtxmath}
\usepackage{amsmath}
\usepackage{array}
\usepackage{balance}
\usepackage{color}
\usepackage{graphicx, floatrow}
\usepackage{subfig}
\usepackage{booktabs}
\usepackage{multirow}
\usepackage{bm}
\usepackage[hyphens]{url}
\usepackage{float}
\floatstyle{plaintop}
\restylefloat{table}
\usepackage{etoolbox}
\makeatletter
\patchcmd{\@makecaption}
{\scshape}
{}
{}
{}
\makeatother
\usepackage{caption}
\usepackage{cleveref}
\begin{document}

\title{AI based Service Management for 6G Green Communications}
\author{Bomin Mao,~\IEEEmembership{Member,~IEEE},
Fengxiao Tang,~\IEEEmembership{Member,~IEEE,}\\
Yuichi Kawamoto,~\IEEEmembership{Member,~IEEE,}
and Nei Kato,~\IEEEmembership{Fellow,~IEEE}

\thanks{Bomin Mao, Fengxiao Tang, Yuichi Kawamoto, and Nei Kato are with the Graduate School of Information Sciences, Tohoku University, Sendai, Japan. 
Emails: 
			\{bomin.mao, fengxiao.tang, youpsan, and kato\}@it.is.tohoku.ac.jp}}

\markboth{}%
{Shell \MakeLowercase{\textit{et al.}}: Bare Demo of IEEEtran.cls for IEEE Communications Society Journals}

\maketitle

\begin{abstract}
Green communications have always been a target for the information industry to alleviate energy overhead and reduce fossil fuel usage. In current 5G and future 6G era, there is no doubt that the volume of network infrastructure and the number of connected terminals will keep exponentially increasing, which results in the surging energy cost. It becomes growing important and urgent to drive the development of green communications. However, 6G will inevitably have increasingly stringent and diversified requirements for Quality of Service (QoS), security, flexibility, and even intelligence, all of which challenge the improvement of energy efficiency. Moreover, the dynamic energy harvesting process, which will be adopted widely in 6G, further complicates the power control and network management. To address these challenges and reduce human intervene, Artificial Intelligence (AI) has been widely recognized and acknowledged as the only solution. Academia and industry have conducted extensive research to alleviate energy demand, improve energy efficiency, and manage energy harvesting in various communication scenarios. In this paper, we present the main considerations for green communications and survey the related research on AI-based green communications. We focus on how AI techniques are adopted to manage the network and improve energy harvesting toward the green era. We analyze how state-of-the-art Machine Learning (ML) and Deep Learning (DL) techniques can cooperate with conventional AI methods and mathematical models to reduce the algorithm complexity and optimize the accuracy rate to accelerate the applications in 6G. Finally, we discuss the existing problems and envision the challenges for these emerging techniques in 6G.  
\end{abstract}


\IEEEpeerreviewmaketitle

\section{Introduction}
\label{intro}
\begin{figure*}[t]
	\centering 
	\subfloat[Energy consumption of ICT and its share.]{\label{energy-consumption-1}  \includegraphics[scale=0.45]{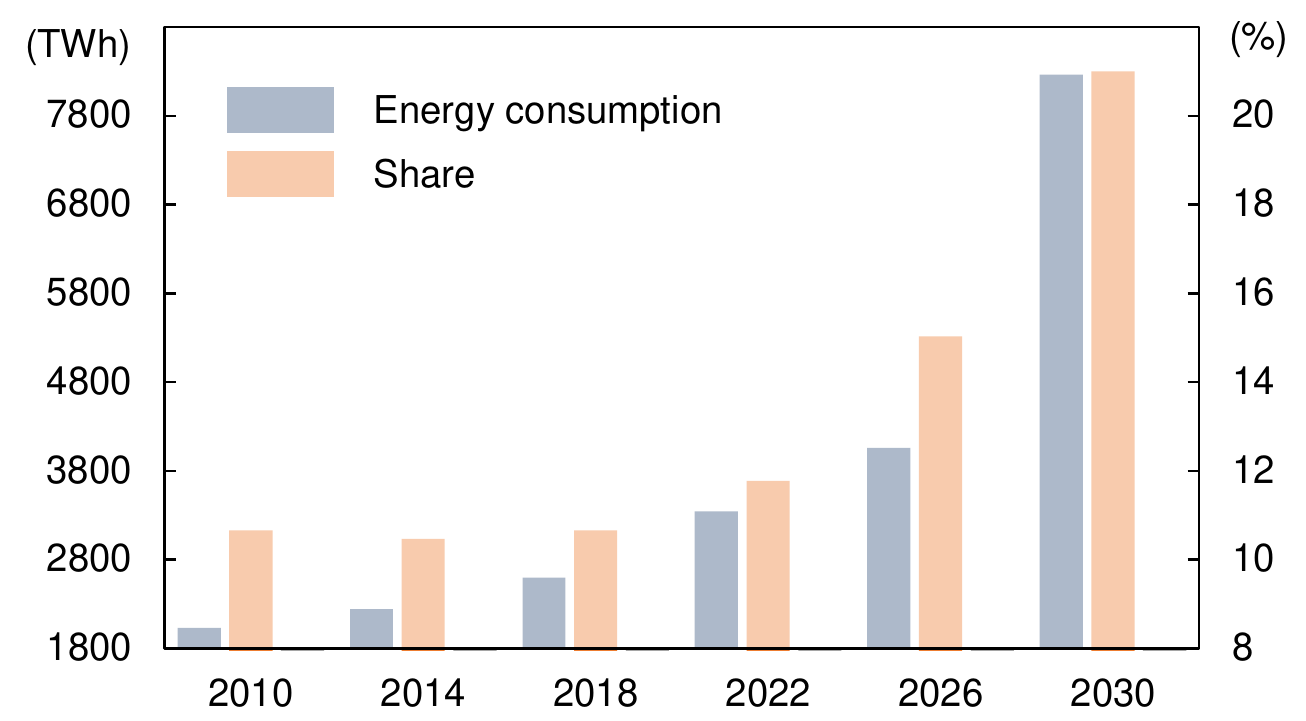}} \hspace{3ex} 
	\subfloat[Energy consumption of different parts for ICT.]{\label{energy-consumption-2}  \includegraphics[scale=0.45]{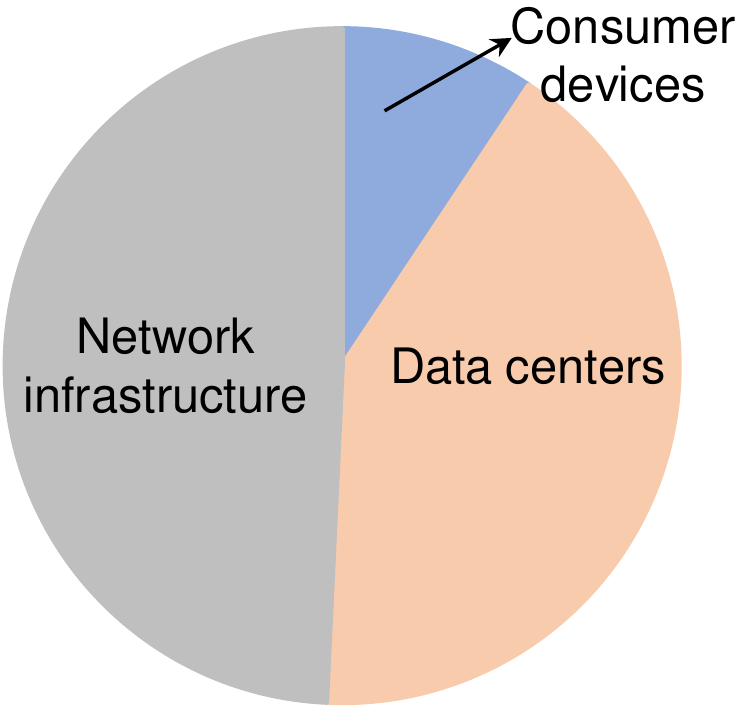}} \hspace{3ex}
	\subfloat[Various energy harvesting sources for ICT.]{\label{energy-consumption-3}  \includegraphics[scale=0.35]{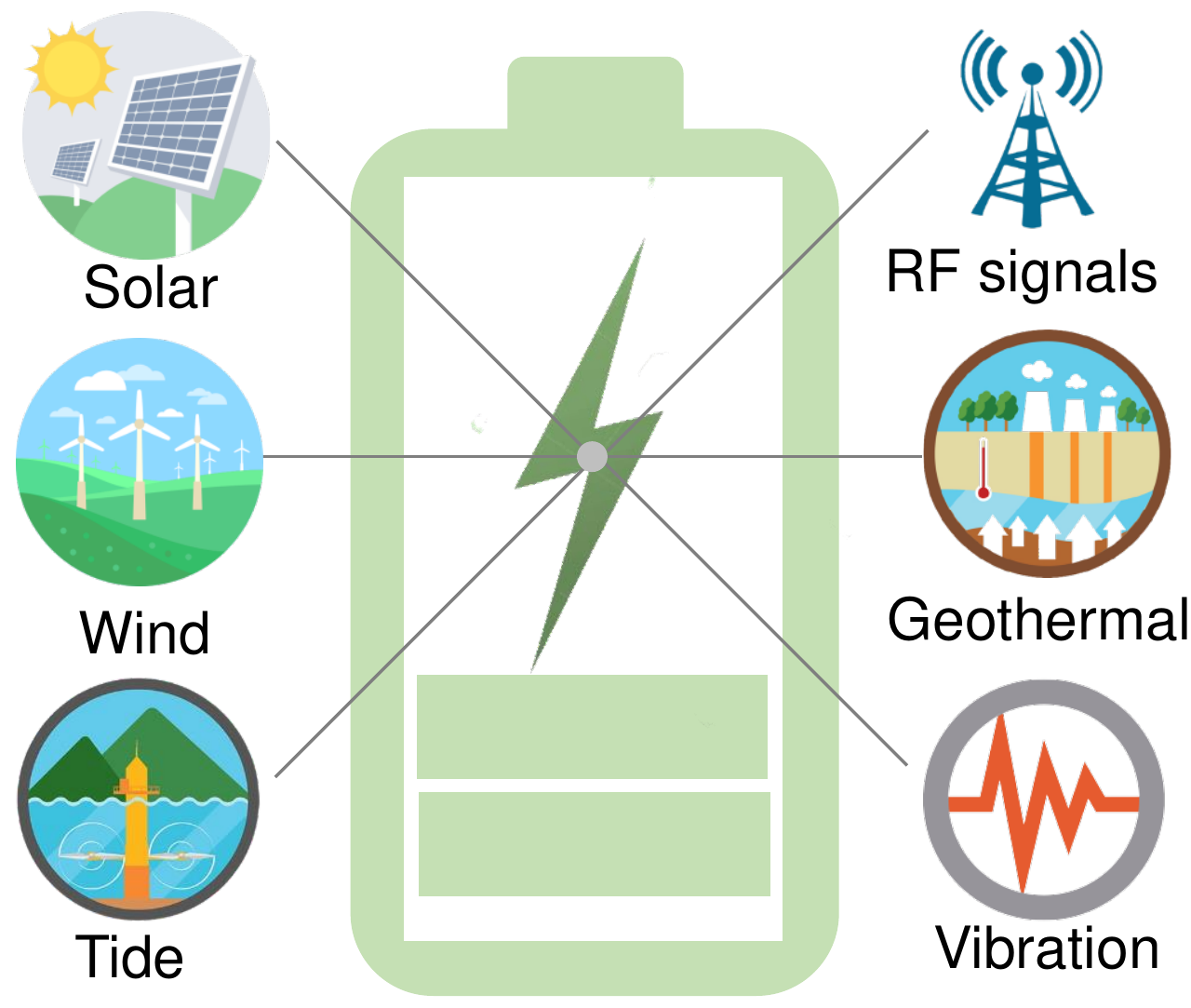}} \hspace{3ex}	
	\caption{Tendency of energy consumption for ICT and the promising energy harvesting techniques.}
	\label{energy-consumption}
\end{figure*}

\IEEEPARstart{R}{ecently}, 5G has been launched to provide users with high-throughput services in some countries, while the worldwide researchers have started to conceive 6G~\cite{khaled19roadmap,david18vision,saad20vision}. It has been reported that 5G Base Stations (BSs) and mobile devices consume much more energy than 4G~\cite{5g-power-whitepaper}. For example, a typical 5G BS with multiple bands has a power consumption of more than 11,000W, while a 4G BS costs less than 7,000W. The dramatically increased power consumption mainly comes from two parts: the growing Power Amplification (PA) in the massive Multiple Input Multiple Output (MIMO) antenna and the processing of booming data. Even though the energy consumption per unit of data has dropped drastically, the exponentially increasing energy required to provide seamless 5G services cannot be neglected since the number of required 5G BSs is at least 4 times that of 4G to cover the same sized area. Data show that Information and Communication Technology (ICT) accounts for more than total electricity consumption as shown in Fig.~\ref{energy-consumption-1}, and it will keep an estimated annual growth rate between 6\% and 9\%~\cite{ict-energy-consumption,andrae15global}.

Then, what will be the situation for 6G in terms of energy consumption? As we know, 6G is expected to extend the utilized frequency bands to Terahertz (THz) for 1,000 times of throughput improvement on the basis of 5G~\cite{khaled19roadmap}. Since the upper bound of transmission range is shortened from 100 m of millimeter Wave (mmWave) to 10 m of THz spectrum, future THz-enabled BS is envisioned to be deployed in the house to provide indoor communications~\cite{6g-whitepaper}, which means significant growth of required BSs. Moreover, besides the communication purpose for mobile terminals and various sensing devices, the computation and content provision services will be gradually transferred from local devices to clouds and edge servers through real-time communications~\cite{lin15time,lin18three}, which is one of the main constituent of ICT energy consumption as in Fig.~\ref{energy-consumption-2}. Another critical paradigm is the utilization of Artificial Intelligence (AI) techniques to provide context-aware information transmissions and personal-customized services, as well as realize the automatic network management~\cite{khaled19roadmap,tang18remove, tang18intelligent}. The growing ICT infrastructure, exploding data, and the increasingly complex network management will result in surging energy consumption, which poses a great challenge for the network operators~\cite{kar18energy,wang20cost}. Data analysis shows that the ICT sector may cost more than 20\% of the total electricity~\cite{ict-energy-consumption} as in Fig.~\ref{energy-consumption-1}. 

\begin{figure}[!t]
	\includegraphics[scale=0.7]{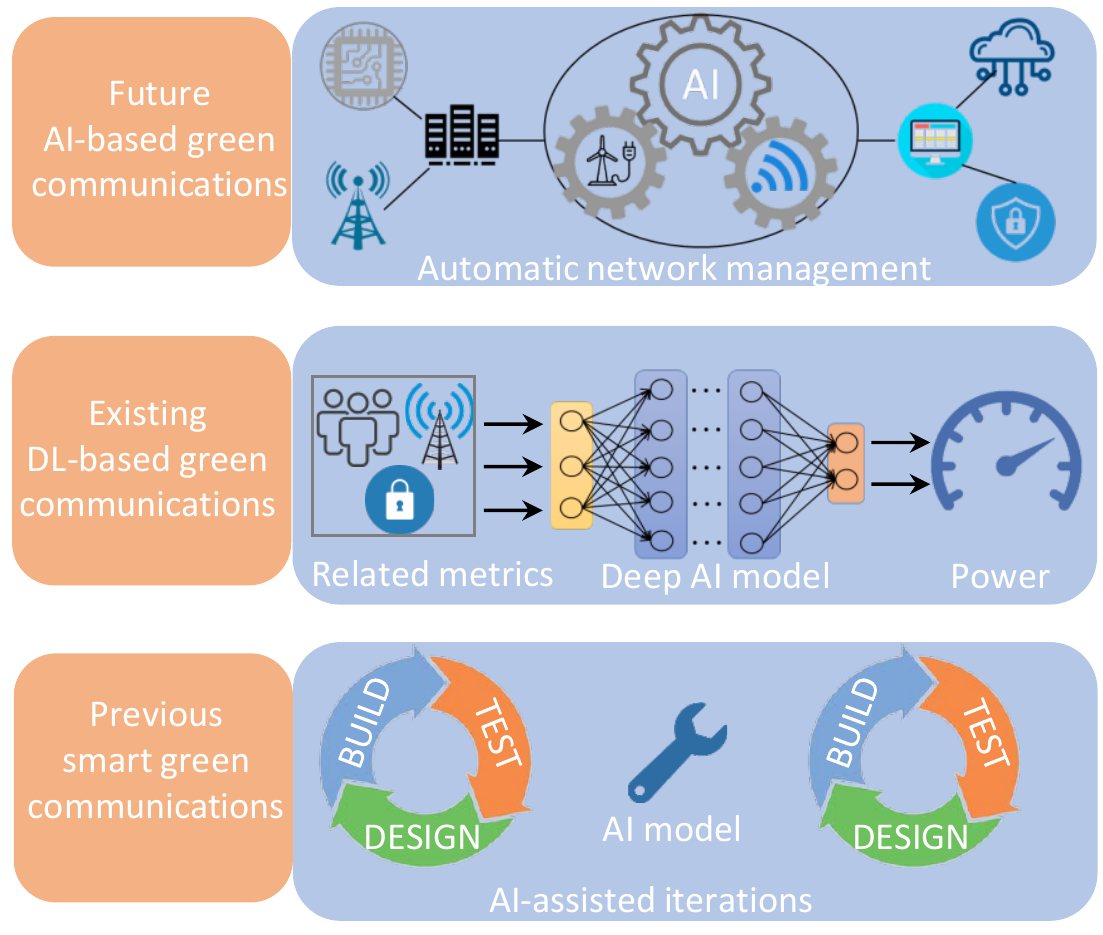}
	\caption{The development of AI-based green communications}
	\label{development-ai-green}
\end{figure}

To alleviate the growing energy burden toward 6G, the academia and industry have conducted extensive research. And the available solutions to address the huge energy consumption mainly come from two parts: energy-efficient network design~\cite{bashar20exploit,vallero19green} and energy harvesting~\cite{mao19harvest,chu19reinforcement}. Specifically, energy harvesting units, such as the solar panels, wind turbines, and vibration harvester, are widely adopted to convert the various kinds of energy to electricity for the communication devices as shown in Fig.~\ref{energy-consumption-3}. Among these energy harvesting techniques, Radio Frequency (RF) harvesting is an important technique which enables not only the simultaneous information and energy transmission, but also the utilization of the interference signal. Similar to RF harvesting, the Intelligent Reflecting Surface (IRS) is expected to be widely deployed to reflect the wasted signal to the receivers to increase the Signal to Interference plus Noise Ratio (SINR)~\cite{hashida20intelligent,bjornson20reconfigurable,dampahalage20intelligent}. Some other deployment including the satellites and Unmanned Aerial Vehicles (UAVs), are deployed to provide seamless coverage. For more efficient energy/power management, AI techniques including conventional heuristic algorithms, the popular Machine Learning (ML), and state-of-the-art Deep Learning (DL) methods, has been adopted to simplify the traditional mathematical iteration process and predict the future network changes as shown in Fig.~\ref{development-ai-green}. Since the future network services have diverse requirements instead of only the high throughput, traditional mathematical models aiming at improving the bit-per-Joule may not be applied to future complex scenarios. To realize the automatic network management toward the green era, AI is the most promising solution. And what we need to do is analyze the various network resources and consider more joint optimizations as shown in Fig.~\ref{development-ai-green}. Accordingly, AI techniques are more widely adopted to optimize the power control and resource allocation in many works~\cite{xiao20reinforcement,zhang20deep2,gong20deep,ren19federate}. In this research, we conduct a survey on AI-related service management for 6G green communications. In the following paragraphs, we introduce the motivations, scope, and contributions of this paper.

\begin{table*}[!ht]
\begin{tabular}{|l|l|l|}
\hline
Publication                                                   & Topics in this survey                                                                                                          & Difference and enhancements of our survey                                                                                                                     \\ \hline
Zhang, 2010~\cite{zhang10energy}               & Energy efficiency, optical networks                                                                                            & \begin{tabular}[c]{@{}l@{}}Focus on energy-efficient wireless \\ communications and network management\end{tabular}                                            \\ \hline
Sudevalayam, 2011~\cite{sudevalayam11energy}   & \begin{tabular}[c]{@{}l@{}}Energy harvesting, wireless\\ sensor networks\end{tabular}                                          & \begin{tabular}[c]{@{}l@{}}Enhanced coverage including various wireless\\scenarios and AI-based green communications\end{tabular}          \\ \hline
Feng, 2013~\cite{feng13survey}                 & \begin{tabular}[c]{@{}l@{}}Energy efficiency, resource management,\\ cooperative communication, MIMO, OFDMA\end{tabular}       &\begin{tabular}[c]{@{}l@{}} Focus on AI-based energy-efficient\\network management \end{tabular}                                                                                                         \\ \hline
Aziz, 2013~\cite{aziz13survey}                 & \begin{tabular}[c]{@{}l@{}}Energy efficiency, wireless sensor\\ networks, topology control\end{tabular}                        & \begin{tabular}[c]{@{}l@{}}Enhanced coverage including various wireless\\scenarios and AI-based green communications\end{tabular}          \\ \hline
Budzisz, 2014~\cite{budzisz14dynamic}          & \begin{tabular}[c]{@{}l@{}}Energy efficiency, cellular\\ networks, WLAN, sleep modes\end{tabular}                              & \begin{tabular}[c]{@{}l@{}}Enhanced coverage of energy-efficient wireless\\ communication scenarios\end{tabular}                              \\ \hline
Lu, 2015~\cite{lu15wireless}                   & \begin{tabular}[c]{@{}l@{}}RF Energy harvesting, SWIPT,\\ CRN, communication protocols\end{tabular}                            & \begin{tabular}[c]{@{}l@{}}Enhanced coverage of various wireless scenarios\\ and focus on AI-based green communications\end{tabular} \\ \hline
Ismail, 2015~\cite{ismail15survey}             & \begin{tabular}[c]{@{}l@{}}Energy efficiency, cellular networks,\\power consumption modeling\end{tabular}                     & \begin{tabular}[c]{@{}l@{}}Enhanced coverage of various wireless scenarios\\ and focus on AI-based green communications\end{tabular} \\ \hline
Fang, 2015~\cite{fang15survey}                 & \begin{tabular}[c]{@{}l@{}}Energy efficiency, information-centric\\ networking, content delivery networks\end{tabular}         & Focus on wireless communication scenarios                                                                                                                     \\ \hline
Erol-Kantarci, 2015~\cite{kantarci15energy}    & \begin{tabular}[c]{@{}l@{}}Smart grid, data centers,\\ energy-efficient communications\end{tabular}                            & \begin{tabular}[c]{@{}l@{}}Focus on energy-efficient wireless\\communications and network management\end{tabular}                                            \\ \hline
Huang, 2015~\cite{huang15green}                & \begin{tabular}[c]{@{}l@{}}Energy efficiency, energy harvesting,\\ cognitive radio networks\end{tabular}                       & \begin{tabular}[c]{@{}l@{}}Enhanced coverage including various wireless\\scenarios and AI-based green communications\end{tabular}          \\ \hline
Peng, 2015~\cite{peng15recent}                 & \begin{tabular}[c]{@{}l@{}}Interference control, energy harvesting,\\ resource allocation, heterogeneous networks\end{tabular} & \begin{tabular}[c]{@{}l@{}}Focus on AI-based energy-efficient\\communications and network management\end{tabular}                                            \\ \hline
Mahapatra, 2016~\cite{mahapatra16energy}       & \begin{tabular}[c]{@{}l@{}}Energy efficiency, tradeoff,\\ spectrum, routing, scheduling\end{tabular}                           & \begin{tabular}[c]{@{}l@{}}Focus on state-of-the-art power management\\ for network performance optimization\end{tabular}                              \\ \hline
Heddeghem, 2016~\cite{heddeghem16quantitative} & \begin{tabular}[c]{@{}l@{}}Power saving techniques in IP-over-WDM\\ backbone networks\end{tabular}                             & Focus on the different wireless access networks                                                                                                               \\ \hline
Ku, 2016~\cite{ku16advances}                   & \begin{tabular}[c]{@{}l@{}}Energy harvesting, usage protocol,\\ energy scheduling, network design\end{tabular}                 & \begin{tabular}[c]{@{}l@{}}Enhanced coverage of green\\communication techniques   \end{tabular}                                                                                                        \\ \hline
Buzzi, 2016~\cite{buzzi16survey}                   & \begin{tabular}[c]{@{}l@{}}Energy efficiency, 5G, cellular\\ network, energy harvesting\end{tabular}                 & \begin{tabular}[c]{@{}l@{}}Enhanced coverage of green\\communication techniques   \end{tabular}                                                                                                        \\ \hline
Omairi, 2017~\cite{omairi17power}              & \begin{tabular}[c]{@{}l@{}}Energy harvesting, wireless\\ sensor networks\end{tabular}                                          & \begin{tabular}[c]{@{}l@{}}Enhanced coverage including various wireless\\networks and AI-based green communications\end{tabular}   \\ \hline
Zhang, 2017~\cite{zhang17fundamental}          & \begin{tabular}[c]{@{}l@{}}Green communications, tradeoffs,\\ 5G networks\end{tabular}                                         & \begin{tabular}[c]{@{}l@{}}Focus on AI-based power management\\ for network performance optimization\end{tabular}                              \\ \hline
Alsaba, 2018~\cite{alsaba18beamforming}        & \begin{tabular}[c]{@{}l@{}}Energy harvesting, beamforming,\\ SWIPT, physical layer security\end{tabular}                       & \begin{tabular}[c]{@{}l@{}}Enhanced coverage of energy-efficient wireless\\ communications and network management\end{tabular}                      \\ \hline
Perera, 2018~\cite{perera18simultaneous}      & SWIPT, 5G                                                                                                                      &\begin{tabular}[c]{@{}l@{}} Focus on energy-efficient communications\\and network management    \end{tabular}                                                                                           \\ \hline
Chen, 2019~\cite{chen19optimization}           & \begin{tabular}[c]{@{}l@{}}Energy-saving, physical-layer and\\ cross-layer communication coding\end{tabular}                   & \begin{tabular}[c]{@{}l@{}}Enhanced coverage including various AI-based\\ energy-efficient communication techniques\end{tabular}       \\ \hline
Tedeschi, 2020~\cite{tedeschi20security}       & \begin{tabular}[c]{@{}l@{}}Energy harvesting, security,\\ green communications, IoT\end{tabular}                               & \begin{tabular}[c]{@{}l@{}}Focus on AI-based energy-efficient\\communications and network management\end{tabular}                                            \\ \hline
Ma, 2020,~\cite{ma20sensing}                   & \begin{tabular}[c]{@{}l@{}}IoT, energy harvesting, sensing,\\ computing, and communications\end{tabular}                       & \begin{tabular}[c]{@{}l@{}}Enhanced coverage including\\heterogeneous wireless networks  \end{tabular}                                                                                                 \\ \hline
Hu, 2020~\cite{hu20modeling}                   & \begin{tabular}[c]{@{}l@{}}Energy harvesting management,\\ 5G/B5G communication networks\end{tabular}                          & \begin{tabular}[c]{@{}l@{}}Enhanced coverage of energy-efficient\\communications and focus on AI-based solutions\end{tabular}                                \\ \hline
\end{tabular}
\caption{Existing Surveys on Energy Harvesting and Green Communications}
\label{existing-surveys}
\end{table*}

\subsection{Motivation}
\label{motivation}
\subsubsection{Energy-related Issues for Different Network Services}
Similar to the 5G which has defined three kinds of services including the eMBB (enhanced Mobile Broadband), uRLLC (ultra-Reliable and Low-Latency Communications), and mMTC (massive Machine Type Communications), some researchers have also considered service definitions in 6G~\cite{khaled19roadmap}. Among these different service definitions, we expand our introductions from three typical communication scenarios: Cellular Network Communications (CNC), Machine Type Communications (MTC), and Computation Oriented Communications (COC). 
\begin{itemize}
	\item \textit{CNC}: Since the majority of energy consumption for cellular networks comes from the BSs, the related research on green CNC mainly focuses on the deployment and configurations of BSs. To optimize energy efficiency of CNCs, the deployment and work states of the BSs should be carefully analyzed and scheduled. Moreover, for the working BSs, the power control and resource allocation are critical to improving the system throughput with minimum energy consumption. Furthermore, the energy harvesting technology can be also considered to alleviate the grid electricity demand of BSs.
	\item \textit{MTC}: For the MTC devices most of which are battery-constrained and difficult to be charged, to alleviate energy demand can be conducted from the access layer and network layer. The research mainly concentrates on the optimization of network access, routing, and relay. As energy harvesting has been widely regarded as an important technique for future Internet of Things (IoT) networks, how to manage the networks considering energy dynamics is challenging and meaningful.
	\item \textit{COC}: Computation and storage services will be an important part of 6G, which is also energy-aggressive as shown in Fig.~\ref{energy-consumption-2}. For the computation parts, the research to reduce energy consumption mainly analyzes the offloading decision computation resource allocation since each server has a limited capacity. Moreover, the uneven distribution of computation demand requires the optimization of server deployment for the balance of latency and energy consumption. For the Content Delivery Networks (CDNs), the content caching and delivery policies directly affect energy consumption. 
\end{itemize}

\subsubsection{Limitations of Conventional Methods}
\label{limitation}
To alleviate energy demand and improve energy efficiency is usually very complex since it is not only concerned with the power control, but also related to many other factors, such as transmission scheduling, resource allocation, network design, user association, and so on. Thus, the formulated problem considering multiple related factors is non-convex or NP-hard~\cite{dong18energy,matthiesen20global,zhang20deep2}. And the conventional mathematical approach is to iterative search the global optimum result or divide into two or multiple sub-problems and search the sub-optimal point~\cite{yang16energy,jiang16energy}. However, due to the increasing factors necessary to be considered, the solution space is significantly huge, resulting in low convergence or extreme difficulty in finding the global optimum. Moreover, since 6G network services have more diversified requirements for throughput, latency, and reliability than 5G, common mathematical optimization methods focusing on the maximization or minimization of a single metric is not enough. Furthermore, the nonlinear and unclear relationship among multiple parameters necessary to be considered makes the mathematical models difficult to be constructed. Additionally, node mobility and service changes lead to increasing network dynamics, which may result in frequent failures of conventional methods. 
  
\subsubsection{Advantages of AI Methods}
Compared with conventional methods, AI techniques including the traditional heuristic algorithms, ML, and the currently popular DL approaches have significant advantages. AI techniques aim to solve the problems in a naturally intelligent manner~\cite{chang20devil}. Thus, it can try to explore the complex relationship among different network parameters through trial and error~\cite{mao19intelligent}. In current years, the ML/DL methods have been widely used to learn the power control and resource allocation policy~\cite{xiao20reinforcement,matthiesen20global,wang19machine,zhang18reinforcement}, which greatly alleviate the difficulty in manually studying the complex relationships and constructing the mathematical models. Moreover, many AI models can estimate the changes of network parameters, which enables the necessary network adjustment in advance and avoids the potential performance deterioration~\cite{zhou18deep,kato17deep}. More importantly, the future increasing Internet users and growing traffic provide massive data resource to adopt and develop AI methods in order to realize automatic network management. 

\subsection{Scope}
\label{scope}
In this paper, we focus on AI-based research to alleviate energy cost and improve energy efficiency. Different from previous works which concentrate on some definite networks~\cite{zhang10energy,huang15green,buzzi16survey,tedeschi20security}, our research is expanded from three 6G communication services: CNC, MTC, and COC. And we mainly focus on AI techniques utilizing for green communications including the traditional heuristic algorithms, ML, and the state-of-the-art DL. Detail introductions will be given in the following paragraphs.

\subsubsection{Existing Surveys}
\label{existing-survey}
The green communications-related topics have attracted scholars' attention in more than 10 years and Table~\ref{existing-surveys} lists the concerned survey papers. We can find that these survey papers focus on definite networks, including backbone networks~\cite{heddeghem16quantitative}, optical networks~\cite{zhang19wireless}, cellular networks~\cite{feng13survey, budzisz14dynamic, ismail15survey, zhang17fundamental, alsaba18beamforming, perera18simultaneous, hu20modeling}, Cognitive Radio Networks (CRNs)~\cite{lu15wireless,huang15green}, and Wireless Sensor Networks (WSNs)~\cite{sudevalayam11energy,aziz13survey, omairi17power}. And different topics, such as improving energy efficiency~\cite{zhang10energy,feng13survey, aziz13survey, ismail15survey, fang15survey, kantarci15energy, mahapatra16energy}, energy harvesting~\cite{sudevalayam11energy, lu15wireless, huang15green, peng15recent, ku16advances, omairi17power, alsaba18beamforming, tedeschi20security, ma20sensing,hu20modeling}, balancing energy cost and network performance tradeoff~\cite{mahapatra16energy,zhang17fundamental} have been discussed. However, no research focus on AI-based energy-efficient communication techniques, even though AI has been regarded as the next paradigm to improve communication and network performance~\cite{khaled19roadmap,kato20ten}. Another problem is that these surveys mainly focus on the relationship between energy and communication performance. However, computation and storage services will be an important part for 6G~\cite{rodrigues20machine, wp20edge}. Thus, to construct the 6G green ICT systems, we need to make analysis from not only the communication perspective, but also the computation perspective.

\begin{figure*}[!t]
	\includegraphics[scale=0.7]{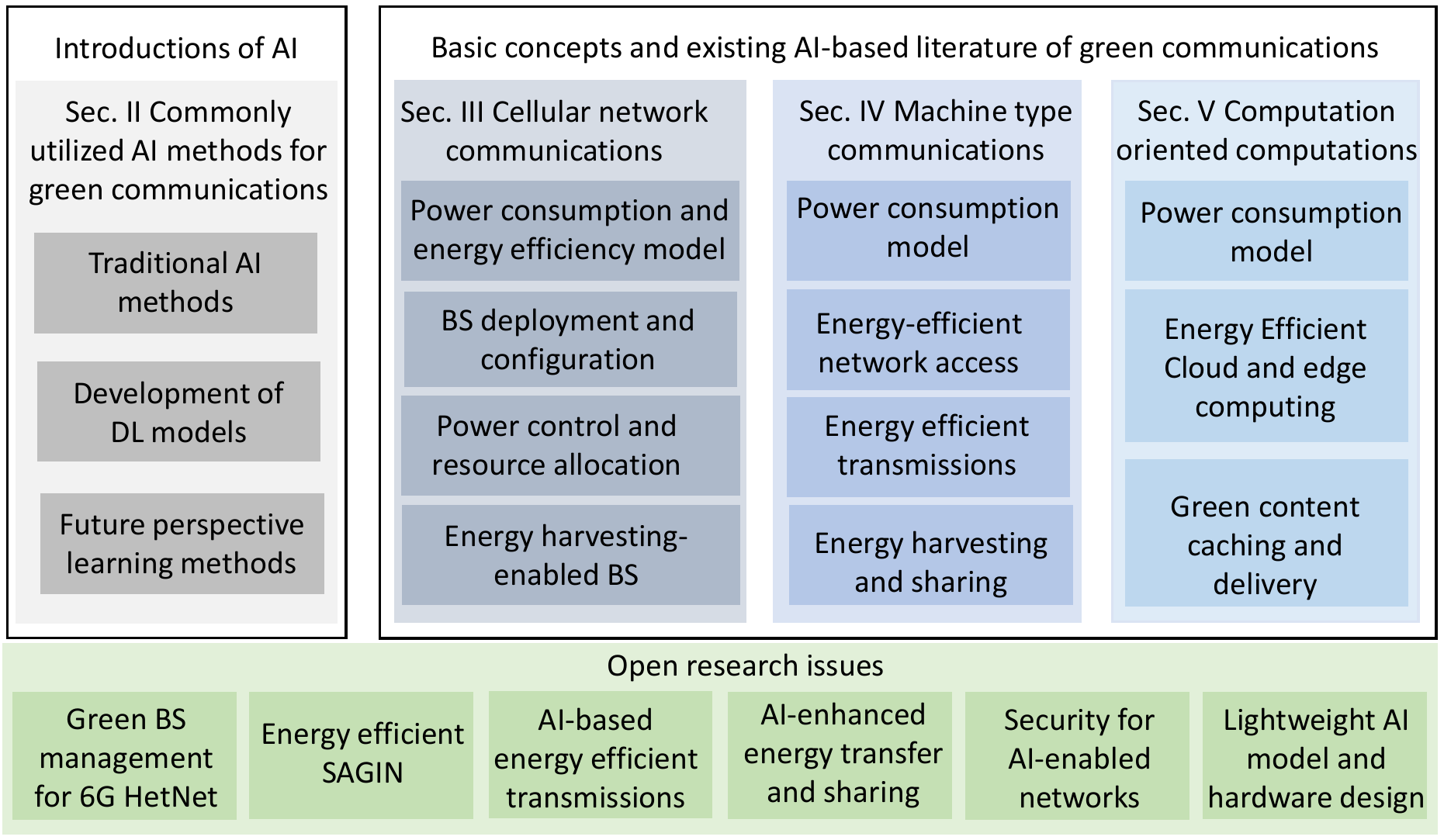}
	\caption{Main contents and structure of this article}
	\label{paper-structure}
\end{figure*}

\subsubsection{Structure of This Survey}
\label{structure}
The remaining part consists of five sections. Before introducing the related research, we introduce the widely-adopted AI techniques in Sec.~\ref{ai-section}. Then, we introduce the related research according to the studied communication scenarios including CNC, MTC, and COC in Sec.~\ref{cnc-section},~\ref{mtc-section}, and~\ref{coc-section}, respectively. Then, we summarize the limitations of existing research and envision the future directions in Sec.~\ref{open-research-section} and conclude this article in Sec.~\ref{conclusion}. The structure of this paper is given in Fig.~\ref{paper-structure}.
\subsubsection{Contribution}
\label{contribution}
After discussing the existing surveys and introducing our research, the contributions can be summarized as below:
\begin{itemize}
	\item We summarize the commonly concerned communication parts and techniques to alleviate energy demand and improve energy efficiency.
	\item We introduce the widely-adopted AI models as well as the state-of-the-art ML/DL methods to improve energy management and network performance, which can give some ideas for future related research.
	\item We analyze the green ICT systems from not only the communication perspective, but also the viewpoint of computation. And this survey covers the most promising 6G network scenarios, including THz-enabled cellular networks, Satellite-Air-Ground Integrated Networks (SAGINs), DCNs, Vehicular ad hoc Networks (VANETs), and IoTs. 
	\item We not only focus on how AI is adopted in these research works, but also analyze how to design AI models to improve the performance. Especially, we explain the common techniques and mathematical methods to improve the AI accuracy rate.
	\item We envision the challenges of AI-based 6G green communications including the overwhelming computation overhead, security issues, and practical deployment. 
\end{itemize}

\section{Overview of AI Methods Towards 6G Energy-Efficient Communications}
\label{ai-section}
Besides the applications in image classification~\cite{li21similar}, natural language processing~\cite{boulianne20study}, and game~\cite{silver17mastering}, AI techniques have been widely studied to optimize the network performance~\cite{mao17tensor,liu18energy,moon16energy,zhao20light}, while green communication is an important application. To improve the performance of AI strategies, various AI models have been developed and some new tendency has appeared toward more intelligent communication management. In this section, we give some introductions about traditional and current AI methods.

AI has been confirmed as an important paradigm for 6G to realize the network automatic management~\cite{khaled19roadmap,kato20ten}. However, the growing network complexity and increasingly stringent service requirements cause great challenges for existing AI techniques. Future intelligent network management depends on the cooperation of various parts: network design, deployment, resource allocation, and so on. To realize the intelligence in every part, various kinds of AI techniques will be adopted.

\subsection{Traditional AI Algorithms}
The development of AI technology can be separated into several stages and Fig.~\ref{ai-development} gives an example. As shown in this figure, the traditional AI techniques utilized in communication networks mainly consist of two types: the heuristic algorithms and ML methods~\cite{beheshti13review}. Even though some ML methods also belong to the heuristic algorithms, such as the Artificial Neural Networks (ANNs) and Support Vector Machines (SVMs), we only consider the non-data-based heuristic models for clear explanations. Thus, the former one mainly utilizes the online search of optimum solution through iterations, while the latter group constructs and train definite models with extensive data to accumulate experience. The following paragraphs will give some more detailed discussions.

\subsubsection{Heuristic Algorithms}
\begin{figure}[!t]
	\includegraphics[scale=0.35]{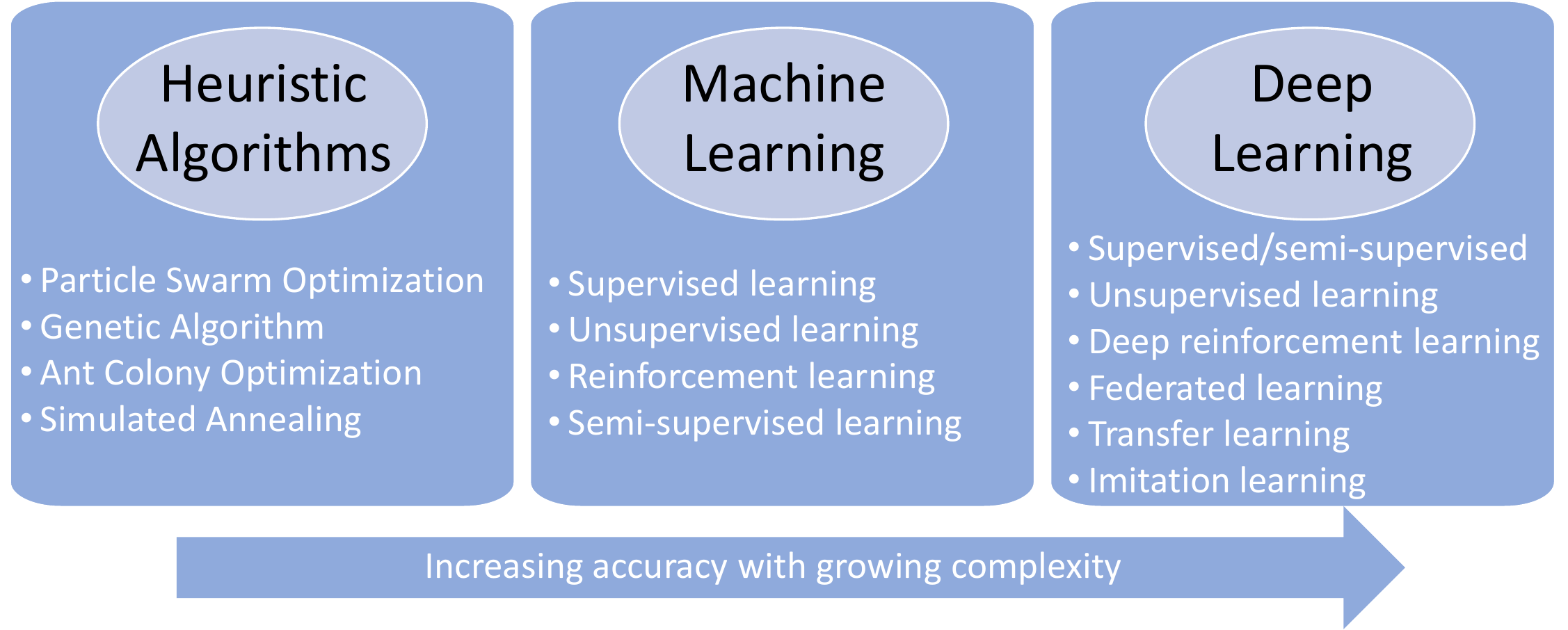}
	\caption{Development of AI Techniques}
	\label{ai-development}
\end{figure}

The heuristic algorithms focus on the NP-hard problem and aim to find a good enough solution given a limited time frame. Generally, the heuristic algorithms use some shortcuts and run faster compared with traditional greedy search methods. However, the sacrifice is the worse accuracy rate or near-global optimum. The shortcut methods vary from different heuristic methods, including the Particle Swarm Optimization (PSO), Ant Colony Optimization (ACO), and Genetic Algorithm (GA) as shown in Fig.~\ref{ai-development}.

\textit{Particle Swarm Optimization}: This optimization method assumes the dubbed particles move around the search-space according to the mathematical formulations of their positions and velocities~\cite{kennedy95particle}. The movement of each particle is affected by its own best position and the best-known positions in the search-space, which leads to the discovery of improved positions. Through repeating the process, a satisfactory solution may be found. This method has been adopted to optimize edge server deployment~\cite{li18energy} and virtual machine placement~\cite{wang13particle,ibrahim20power} in order to improve energy efficiency. Moreover, the method previously mainly adopted for continuous problems has also been illustrated its availability for a discrete process~\cite{wang13particle,ibrahim20power}. However, this method is easy to fall into local optimum in high-dimensional space and has low a convergence rate. 

\textit{Ant Colony Optimization}: Inspired by the ants' behavior to search food, ACO has been proposed to find the optimal route through simulating the revolution~\cite{dorigo06ant}. Similar to PSO, ACO is also based on swarm intelligence, where a grout of artificial "ants" which are multiple simulation ants move through the search space to find the optimal route. And for each artificial ant, record its position and quality, which can guide other ants to locate better positions in later simulation iterations. This method has been widely studied in many network applications in order to improve energy efficiency, such as routing~\cite{li15energy}, resource allocation~\cite{liao17ant}, and server deployment~\cite{liu18energy}. 

\textit{Genetic Algorithm}: The GA, which is also termed genetic programming, borrows the concepts of mutation, crossover, and selection in evolutionary biology to improve the solution~\cite{mallawaarachchi17introduction}. In GA, a group of candidate solutions is abstracted as chromosomes or phenotypes and a pair of chromosomes or phenotypes can crossover to generate a new generation with a certain probability. Moreover, the mutation may happen for each new generation to result in a totally new chromosome or phenotype. To guide the process toward the expected direction, fitness is defined to evaluate the individuals in every generation and the individual with low fitness value is eliminated. GA is easy to converge and expandable, while it cannot guarantee the global optimum and depends heavily on the parameter selection. Researchers have adopted this method to design the cellular networks~\cite{dai20propagation,moysen16machine} and optimize the edge server deployment~\cite{zhang19genetic,gong18set}.

\subsubsection{Machine Learning Algorithms}
As a data-based technique, various ML algorithms have been developed and adopted in many network performance optimization strategies~\cite{fadlullah17state,rodrigues20machine,keller85fuzzy}. In this part, we focus on three machine learning algorithms: regression analysis~\cite{chatterjee15regression}, SVM~\cite{svm-introduction}, and K-means clustering~\cite{k-means-introduction}, which are commonly utilized in green communications. Another important technique: Reinforcement Learning (RL) will be introduced in the next subsection. 

\textit{Regression Analysis}: This method is mainly utilized to analyze the relationship between two or among multiple parameters. The most common application is to map from the input parameters to the output results with the labeled dataset and a cost function is usually defined to evaluate the accuracy rate. According to whether the output is linear or binary, the regression analysis can be divided into linear regression and logistic regression. Regression analysis plays an important role in green communications. For instance, the linear regression can be utilized to predict future traffic changes, which is further adopted to determine the energy-efficient transmission schemes, resource allocation, and computation offloading~\cite{vallero19green}.

\textit{Support Vector Machine}: SVM is adopted to analyze data for classification and regression analysis in a supervised learning manner~\cite{svm-introduction}. An SVM utilizes a set of orthogonal vectors to define a hyperplane or a set of hyperplanes to separate the training data point. And the best hyperplane is the one that has the largest distance to the nearest training data in any class. The SVM can be adopted for high-dimensional problems and suitable for the small dataset. In green communication management, SVM has been applied to solve the problems like user association~\cite{moon16energy} and \textcolor{black}{computation offloading~\cite{wang2020federated}.}

\textit{K-means Clustering}: This method aims to partition multiple observations into several clusters in which each observation belongs to the cluster with the nearest center~\cite{k-means-introduction}. As an unsupervised learning method, this technique repeats the process to assign the nodes into different clusters and update the cluster center. To evaluate the assignments, a cost function based on the distance between the nodes and the cluster center is defined. K-means clustering is efficient to cluster the users and associate them to suitable BSs for saving energy~\cite{visalakshi09kmeans,zhang20energy,zhang20energy2}. It can also be applied to the optimization of cloudlet placement~\cite{shen19energy}.

\subsection{Development of Deep Learning Models}
Since the common ML/DL models and three training manners shown in Fig.~\ref{ai-development} have been introduced in many works~\cite{fadlullah17state, mao18novel}, we just give some discussion about the development of ML/DL models which have been utilized to improve energy efficiency. 

Most of the current ML/DL models are developed from Artificial Neural Networks (ANNs) which can be also termed Neural Networks (NNs). ANN is constructed by layers of interconnected units named "artificial neurons", which is to model the neurons in a biological brain~\cite{krogh08artificial}. Each artificial neuron can process the received signals with some non-linear functions and then transmit the result to neurons in the next layer through the weighted edges. Thus, the final output of each ANN depends on not only the input signals, but also the utilized non-linear functions and edge weights. In recent decades, the ML/DL models have developed fast on the basis of ANNs, which can be summarized into three aspects. First, the most obvious development is the increased number of layers, which result in the deep architectures from traditional shallow ones. Thanks to the breakthrough in the training algorithm~\cite{hinton06fast} as well as the hardware developments, current DL models can have very complex architectures while keeping a extremely high accuracy rate, which enables them to be adopted in very complicated scenarios and overwhelm humans in some applications, such as the board game~\cite{silver17mastering}. Second, connection manners become more complex. Besides the full connections among neurons in adjacent layers for most ANNs, the partial connections have also been utilized in some modern ANNs, such as the Convolutional Neural Networks (CNNs)~\cite{krizhevsky17imagenet}, which enables the flexible processing of the input where features are not distributed everywhere. 
And part of the output can be also further input the learning models, such as the Recurrent Neural Networks (RNN)~\cite{mikolov11extensions}, to generate the time-consecutive variables. Third, researchers have developed the models to concurrently utilize multiple ANNs to cooperatively complete one task, such as the Generative Adversarial Network (GAN)~\cite{goodfellow14generative} and Actor-Critic (AC) method~\cite{konda00actor}. The two ANNs can have the same or different structures while act different roles. Forth, the techniques such as the different activation functions, data processing methods, and attention mechanism significantly improve the accuracy rate of current ML/DL structures.

\subsection{Future Perspective AI Learning Methods}
Besides the development in the ML/DL structures, the learning methods also critically affect the accuracy rate and computation performance. Future networks will consist of more complex scenarios and dynamics, which drives us to consider more advanced AI learning methods. In this part, besides the traditional supervised learning and unsupervised learning, we focus on three AI learning methods which will definitely attract more attention as shown in Fig.~\ref{ai-development}.

\subsubsection{Deep Reinforcement Learning}
RL is the dynamically learning through trial and error to maximize the outcome. In an RL model, the essential components are the environment, a defined agent, the state space, the action space, and reward~\cite{lavet18introduction}. In the studied environment, the agent chooses an action according to the current state, and then gets rewarded for the correct action or penalized for an incorrect one. In the training process, the agent follows the existing experience or explores a new action with a certain probability in order to maximize the reward. In the traditional RL model, a table is usually utilized to store the Q value which is the expected accumulated reward for different actions at each state. The training process is to fill in the table, which can guide future action selection. However, with the studied problem becoming complex, the number of states and potential actions will be huge and even unavailable, which makes the Q-value table impossible. To solve this problem, the DL models are adopted to map from the state to the corresponding action, which is the main concept of Deep Reinforcement Learning (DRL)~\cite{lavet18introduction}. Another advantage is that this method enables an agent to generalize the value of states it has never seen before or just has partial information. Due to these advantages, it has been witnessed that DRL has attracted more attention to improving energy efficiency through optimizing the BS management~\cite{temesgene20distribute}, resource allocation~\cite{he19joint,simsek15learning}, power control~\cite{xiao20reinforcement,zhang20deep4}, and computation offloading~\cite{gong20deep,kwok96dynamic,ren19federate}. 

\subsubsection{Transfer Learning}
Transfer learning is a machine learning method which aims to utilize the constructed knowledge system while solving a problem to the different but related problem~\cite{pan10survey}. Different from traditional ML models which learn the knowledge from zero, what is necessary to do for the new application in related problems is fine-tune the new model based on existing knowledge system or train part of it. Thus, transfer learning can significantly reduce the computation consumption and required training data, resulting in extended and accelerated applications. As the network changes frequently due to the mobility and transmission environment changes, transfer learning is widely considered to address the similar scenarios~\cite{sharma17energy,sharma19transfer,dong20deep,pradhan20computation}. On the other hand, the application range of the existing knowledge system as well as the balance between training and performance in target scenario are hot topics and require more attention in existing research~\cite{dong20deep}.

\subsubsection{Federated Learning}
Federated learning is a decentralization method by utilizing the distributed servers or devices to train and test AI models with the local data~\cite{yang19federated,federated-learning-google-comic}. Thus, the edge servers or devices can keep the training data locally and just need to upload the obtained parameters to the central controller. What the central controller needs to do is collect and integrate the parameters of AI models. And then the edge devices can download AI models to make predictions or conduct periodical update. Since personal privacy arouses increasing concern recently, the federated learning technique will attract growing attention in 6G. Moreover, the cooperative training and running manner of federated learning can efficiently utilize the idle computation resource and reduce the consumption in the central controller. Furthermore, the uploading of parameters instead of training data results in reduced communication overhead~\cite{ren19federate,shen20computation,wang2020federated}.

\subsection{Summary}
From the above introduction, we can find AI techniques have various application scenarios and should be chosen according to definite problems. And with the development of computation hardware, DL techniques have attracted growing attention to solving more complex problems. However, this does not mean that the traditional AI techniques such as heuristic algorithms and shallow ML models are not suitable anymore. Since many traditional AI methods have much lower computation complexity compared with DL, they are suitable for some resource-limited scenarios. In the following paper, we give more detailed explanations about how these methods to realize green communications in different scenarios. It should be noted that some important AI techniques are not introduced in this section, but they still have promising perspectives, such as imitation learning~\cite{wang20imitation} and quantum machine learning~\cite{nawaz19quantum}.

\section{Cellular Network Communications}
\label{cnc-section}

\begin{figure}[!t]
	\includegraphics[scale=0.25]{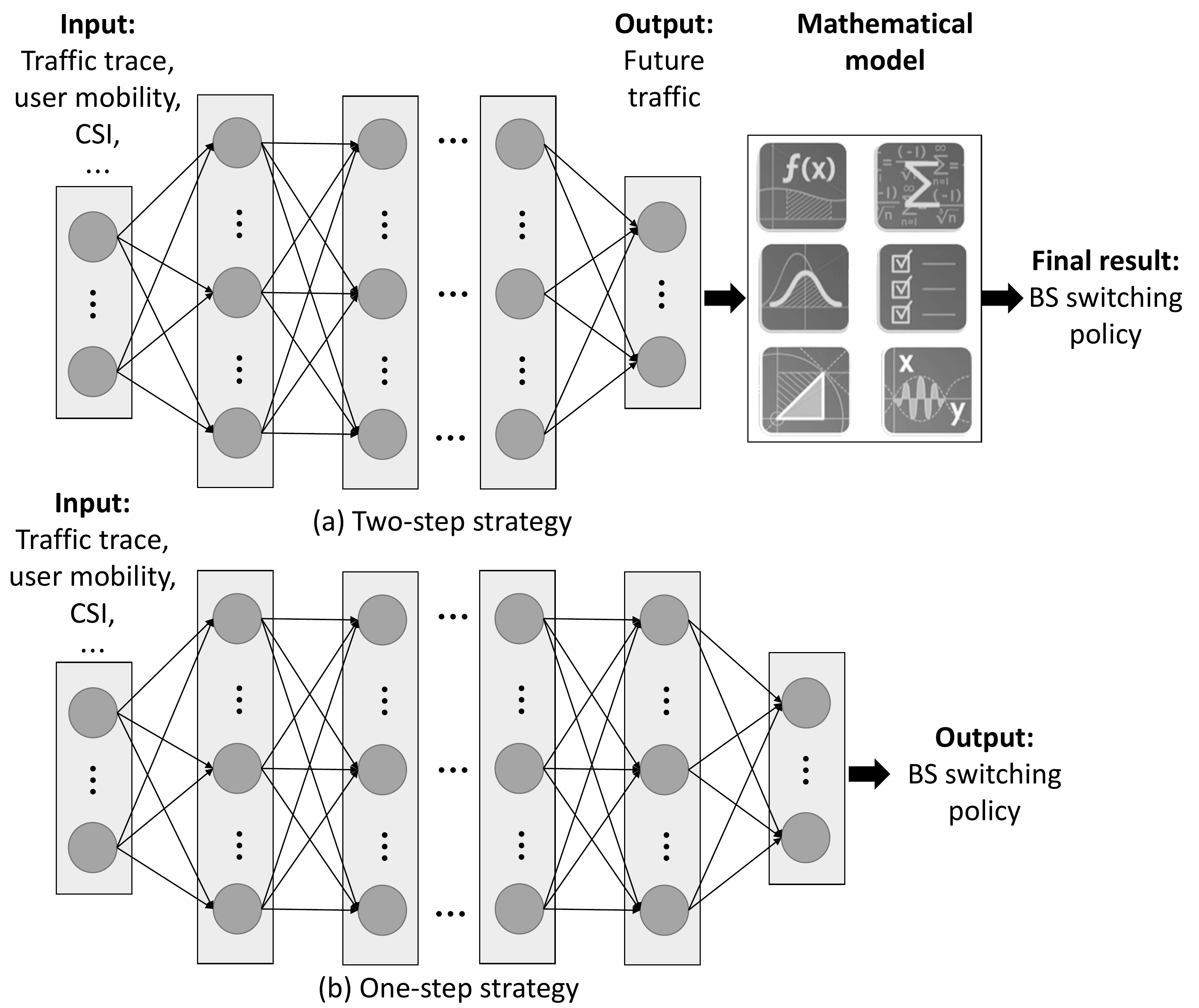}
	\caption{One-step and two-step AI-based BS switching strategy}
	\label{direct-indirect}
\end{figure}

Energy consumption of cellular networks comes from the radio access part and the core part~\cite{ismail15survey}. Some practical measurements of energy consumption of cellular networks have been reported in~\cite{ismail15survey,alsharif2017green}. And the data illustrate that the BSs account for more than half of the total energy consumption, in which more than 50\% to 80\% is utilized for the power amplifier and feeder. With the utilized frequency band extended to sub-THz and THz in the 6G era, the coverage of single BS further shrinks~\cite{khaled19roadmap,fadlullah20heterogeneous}. Then, the required increasing number of BSs to realize seamless coverage is expected to consume more energy. Therefore, green communication research for cellular networks mainly focuses on BSs. In this section, we first introduce the power consumption and energy efficiency modeling of cellular networks and then explain the related AI-based approaches to realize the green communications from different perspectives.

\subsection{Power Consumption and Energy Efficiency of Cellular Networks}
According to our above introductions, we mainly focus on the Radio Access Network (RAN) part consisting of BSs and access terminals. We introduce the power consumption modeling of BSs and the metric "bit-per-Joul" to measure energy efficiency for both the BSs and access terminals.

\subsubsection{Power Consumption Modeling of BSs} The power consumption of a BS consists of four part: power supply, signal processing, air conditioning, and the power amplifier~\cite{ismail15survey}. Since part of the power consumption is constant for BSs at sleep and idle states while the other part is relevant to the workload, energy consumption of a BS can be usually summarized as~\cite{wang19reinforcement}:

\begin{equation}
\label{bs-power}
	P_{bs}=P_{sleep}+I_{bs}\{P_{add}+\eta{P_{trans}}\}
\end{equation} 
\noindent{where $P_{bs}$ and $P_{trans}$ denote the total power consumption and maximum transmission power consumption of the BS, while $\eta\in{[0,1]}$ denotes the usage rate. $P_{sleep}$ is the constant power consumption to sustain the basic functions in sleep mode. $P_{add}$ denotes the additional constant power for computation, backhaul communication, and power supply in active mode. $I_{bs}$ is a binary parameter representing whether the BS is active or sleep.} According to Equation~\ref{bs-power}, to reduce energy consumption, we should try our best to turn the idle BS to sleep mode and minimize the usage in active mode. If we further consider that future deployed multi-tier heterogeneous BSs are enabled with various frequency bands up to THz~\cite{dai20propagation,thakur18energy}, to reduce the consumed energy of all the BSs should be mainly dependent on the BS deployment as well as management, user association, and resource allocation.

\subsubsection{Energy Efficiency Measurement} Energy efficiency is to measure to achieved performance with energy per unit mass. Thus, in the cellular networks, it is usually defined as the ratio between the obtained transmission rate and power consumption with the unit of "bit-per-Joule". Different from the direct energy-saving strategies, to improve energy efficiency is also an important direction towards green communication. Here we deduce the equations of energy efficiency for a UE in cellular networks and then analyze the potential optimization strategies. It should be noted that the derivation method also applies to the BSs.

We assume a multi-cell interference network with multiple single-antenna UEs and several multi-antenna BSs. The same spectrum resource is multiplex among the cells~\cite{matthiesen20global}. If one UE's transmission power and the channel gain to the corresponding BS are $P_{t}^{u}$ and $G_u$, respectively, then the maximum uplink transmission rate can be calculated as:
\begin{equation}
	R_u=B\log{(1+\frac{GP_u}{N+I})}
\label{transmission-rate}
\end{equation} 
\noindent{where $R^u$ is the maximum transmission rate for the uplink of the considered UE. And $B$ is the assigned bandwidth, while $N$ and $I$ denote the noise and interference on the utilized channel, respectively.} If we further assume the inefficiency of the considered UE's power amplifier and static power consumption are $\mu$ and $P_{u0}$, respectively, then energy efficiency can be calculated as below:
\begin{equation}
	EE^u=\frac{R^u}{\mu{P_u}+P_{u0}}
	\label{energy-efficiency-user}
\end{equation}

According to Equations~\ref{transmission-rate} and~\ref{energy-efficiency-user}, we can find that the parameters affecting energy efficiency include the assigned bandwidth, channel gain, transmission power, and interference, while the noise and static power consumption are usually constant. Therefore, we need to optimize the allocation of resource including channels and bandwidth, power control, and transmission scheduling policy to improve energy efficiency.

\subsubsection{Summary} According to our above analysis, the strategies toward the green cellular networks in the 6G era mainly consists of the deployment and management of BSs~\cite{wu20power,alnoman19computing,lai19clustering}, the power control~\cite{xiao20reinforcement,doan20power}, and resource allocation~\cite{wei18user,thakur18energy,zhang19artificial,liu20deep}. Another important direction which has been mentioned in Sec.~\ref{intro} is the utilization of renewable energy to drive the BSs~\cite{miozzo20coordinate,wakaiki19control,ghazanfari16ambient,lin15distribute,kariminezhad19heterogeneous,lin15distribute,kariminezhad19heterogeneous}.

\begin{figure}[!t]
	\includegraphics[scale=0.35]{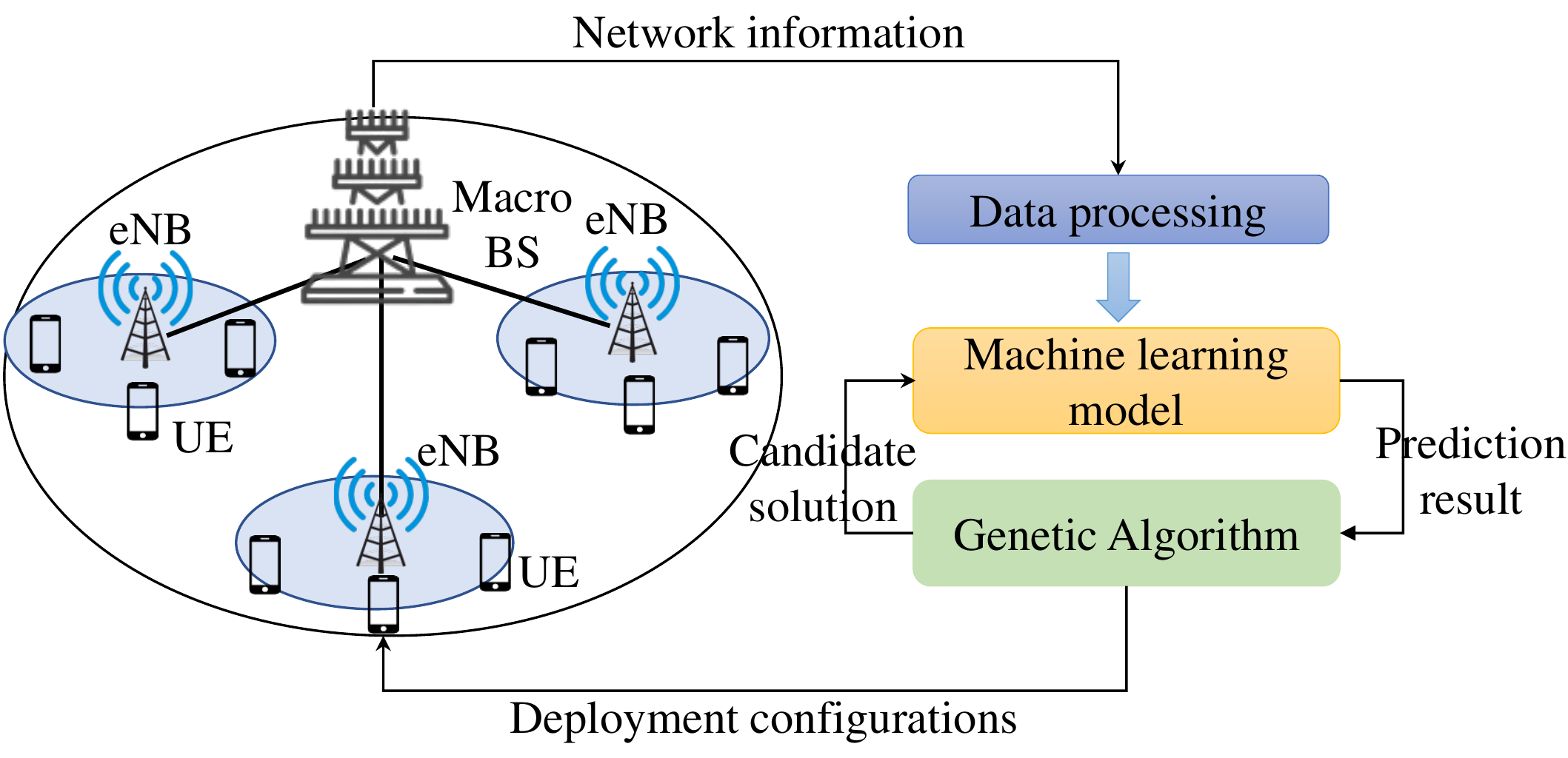}
	\caption{Intelligent BS deployment.}
	\label{intelligent-bs-deployment}
\end{figure}

\subsection{Base Station Deployment and Configuration}
\label{bs}

As we mentioned in Sec.~\ref{intro}, the significant penetration loss of THz radio signals will cause 6G BSs to cover very limited areas with increased available frequency bands~\cite{khaled19roadmap}, both of which contribute to the drastic increase of energy consumption~\cite{5g-power-whitepaper}. However, the uneven distribution and user mobility result in unbalanced traffic load for different BSs. According to Equation~\ref{bs-power}, to reduce energy consumption and improve efficiency, we need to minimize the number and transmit power of working BSs in the cellular networks. Thus, the BS deployment policy, workload management, and user association are three attractive strategies.

\subsubsection{Base Station Deployment}

In the network construction period, the BS deployment is an important factor to affect the communication performance and energy consumption. Even though some deployment positions can be manually selected according to the population density~\cite{borah19effect}, the increasing dynamics, variable propagation characteristics, complex physical surroundings, and even the climates drive the researchers and operators to consider more efficient and automatic strategies. 

To decrease the number of deployed BSs, Dai and Zhang~\cite{dai20propagation} consider multi-objective GA. In their research, the proposed approach firstly extracts the main features which determine the strength of the Received Signal Strength (RSS). Then, multiple ML models including $k$-Nearest Neighbor (KNN)~\cite{keller85fuzzy}, random forest~\cite{random-forest}, SVM~\cite{svm-introduction}, and Multi-Layer Perceptron (MLP)~\cite{marius09multilayer} are adopted to map the relationship between the extracted features and RSS values. In the second stage, the multi-objective GA~\cite{mallawaarachchi17introduction} is adopted to optimize the locations and operating parameters. Specifically, the GA programming process is conducted with a different number of BSs, and then the minimum number reaching the coverage requirement is selected. Then, the feasible solutions are evaluated by the proposed ML models. Simulation results illustrate that the MLP outperforms other ML models in terms of Mean Absolute Error (MAE). And the coverage rate is improved by 18.5\% compared with real-world deployment.

Besides the BS deployment planning, the coverage design is an important factor to affect the required number of BSs and network performance. Assuming the deployment is done without detailed cell planning, Ho. et al~\cite{ho13online} utilize the GA~\cite{mallawaarachchi17introduction} method to adjust the femtocell coverage in order to optimize the three network metrics: coverage holes, coverage leakage, and load balance. In this paper, the authors consider three metrics including coverage holes, coverage leakage, and load to define the fitness function for the evaluation of considered solutions during the evolution process. To overcome unknown network dynamics and user mobility, the online learning method based on periodical updates with real-time network measurements is adopted. In their proposal, the hierarchical Markov Models (hMMs)~\cite{murphy12machine} are used to capture the behavior and generate the load trace of each femtocell with a high accuracy rate. Then, the results can be used to calculate the fitness. And the evolution process is illustrated to provide the continuous performance improvement. 

Similar to~\cite{dai20propagation, ho13online}, Moysen et al.~\cite{moysen16machine} also combine the GA and ML in the design of cellular networks. In their research, the SVM~\cite{svm-introduction} is trained offline as a QoS regressor with the collected data including the Reference Signal Received Power (RSRP) and Reference Signal Received Quality (RSRQ) coming from the serving and neighboring eNBs. Then, in the online phase, the GA algorithm is utilized to generate the feasible solutions consisting of the configuration parameters of eNBs. And then the UE measurements for each feasible solution is utilized as the input of SVM, of which the predicted QoS result is adopted to calculate the fitness function. With the goal of minimizing the PRB per transmitted Mb, the improved BS configuration set can be found through the iterations of GA. The case study illustrates the proposed model can enable the operator to find the appropriate deployment layout and minimize the required resources.

From the above research, it can be found that the deployment policy is usually found by iterative algorithms, such as the GA, while the supervised learning-based training is adopted to predict the multiple network parameters as the input of GA or evaluate the fitness function as shown in Fig.~\ref{intelligent-bs-deployment}. The combinations of the heterogeneous algorithms and ML as shown in Fig.~\ref{intelligent-bs-deployment} can cooperatively improve the performance of the proposed model. Since the DL has shown improved accuracy rate and more advanced policy searching ability, it is highly expected to witness the application of the prevalent DL techniques in the BS deployment design.

\subsubsection{Work State Management}
As the network traffic is dynamically changing due to user mobility, the multi-tier BSs can be scheduled to switch on and off to reduce energy consumption~\cite{feng17base}. If the work state of the BS is changed, the user association information should be adjusted accordingly to ensure a qualified connection. Therefore, the work state of BSs should be scheduled carefully to minimize energy consumption as well as meet the QoS requirement. 

Since the users' daily movements contribute to the similar changing tendency of the traffic patterns, the correlation between the current traffic data and historical experience can be utilized to design the BS switch on/off policy~\cite{sharma17energy,sharma19transfer}. To predict the future traffic with a historical profile and switch off the BSs with low usage may be the easiest solution. The main concern to switch off some BSs is the potential deterioration of QoS. To alleviate the concern, the accuracy rate of traffic prediction affects network performance in terms of energy saving and QoS. Gao et al.~\cite{gao20machine} compare multiple ML models including Auto-regressive Integrated Moving Average (ARIMA)~\cite{arima-wiki}, prophet, random forest, LSTM, and ensemble learning in terms of accuracy rate, speed, and complexity. Then, these models are utilized in traffic prediction. The prediction results are further utilized to calculate energy efficiency. Thus, some BSs can be switched off if the Key Performance Index (KPI) is below the predefined threshold. Similarly, Donevski et al.~\cite{donevski19neural} utilize two kinds of NNs, including the dense NN and RNN to predict the future traffic of Small Base Stations (SBSs) according to the previous trace. Then, a threshold is defined to decide whether the SBS could be switched off or kept on. Another unified strategy is given by directly utilizing the traffic trace to predict the BS switching scheme as shown in Fig.~\ref{direct-indirect}. It should be noted that the threshold in this proposal is adjustable to achieve a balance between the coverage loss and efficiency loss. Simulation results illustrate that energy consumption can be reduced by 63\%, while more than 99.9\% of requests can be satisfied.

Different from the above scenarios which only consider two work states, Pervaiz et al.~\cite{pervaiz18energy} analyze the switching policy for the multi-sleep-level-enabled BSs in a two-tier cellular network. The machine learning technique is utilized to decide the best sleep level of SBSs, while the users keep connections with the Macro Base Stations (MBSs). Specifically, the SVM regression model is considered to predict the vacation period and operation time of the SBSs according to historical network traffic profile. Then, the prediction results are analyzed along with energy consumption and latency to decide which sleep level the SBS should be switched to. It should be noted that the SVM utilized in this paper can be replaced by other regression models.

The above research works utilize the historical traffic profile to efficiently train the ML models in a supervised manner. Researchers have also proposed the approaches to combine the RL and transfer learning to increase the flexibility and accelerate the convergence. Authors of~\cite{sharma17energy,sharma19transfer} consider the RL agent to select the BS work modes for system power minimization according to the traffic patterns. Moreover, transfer learning~\cite{pan10survey} is exploited to use the past learning experience in current scenarios, which can accelerate the learning process. However, these two research works~\cite{sharma17energy,sharma19transfer} neglect the QoS even though the authors consider the user association policy after switching off some BSs. To solve this problem, in~\cite{li14transfer}, the cost function of the RL model is defined as an adjustable combination of energy consumption and service delay instead of only energy consumption~\cite{sharma17energy,sharma19transfer}. Consequently, their proposal can not only reduce energy consumption, but also guarantee the diversified QoS requirements. Additionally, the transfer learning technique is utilized to accelerate the convergence of the considered AC model~\cite{konda00actor}. Another similar research~\cite{zhao14transfer} also combines the RL and transfer learning to design the BS switching policy. In this proposal, the learned knowledge for spectrum assignment is transferred to the process of user association.  

Deep Q-learning (DQL) technique has also been applied to design the BS switching policy based on the network traffic in~\cite{liu18data}. Different from the research~\cite{sharma17energy,sharma19transfer} which directly utilizes the traffic pattern, authors in~\cite{liu18data} consider a traffic modeling module to iteratively fit an Interrupted Possion Process~\cite{fischer93markov} and predict the next traffic belief state. Since the traffic model is learned in an online fashion, it can capture the complex dynamics of real-world traffic, which allows the adopted DQL model to output more accurate action. The adopted Deep Q-network (DQN) decides the sleeping policy according to the output brief state of the traffic modeling module. And the reward function is defined as the sum of the operation cost and the service reward. To enhance the original DQN model, a reply memory storing a certain amount of past experiences are utilized in the training step as a bootstrapped estimation of true distributions. And the stable parameters are stored by a separate network to avoid the training oscillations and divergence. The authors also apply adaptive reward scaling to match the network outputs. Even though the research neglects the mutual effects among BSs, the proposed model is suitable for BSs with different traffic patterns. And the experiment with a network simulator and dataset illustrates the advantages of the proposed model over other ML algorithms.

In the above research, to switch off some BSs in low usage on the one hand reduces energy consumption, on the other hand sacrifices some network performance due to the resulted coverage hole. Therefore, the proposed AI approaches usually define a weighted sum of energy consumption and QoS as the reward or cost function to reach a balance~\cite{pervaiz18energy,li14transfer}. To address the QoS sacrifice physically, Panahi et al.~\cite{panahi18green} consider the heterogeneous scenario where the Device-to-Device (D2D) technique is utilized to relay the messages toward working BSs. To decide the work state for each MBS and Femtocell Base Station (FBS), the authors propose the Fuzzy Q-learning (FQL) algorithm which combines the Q-learning (QL) and Fuzzy Interference System (FIS)~\cite{panahi14optimal,panahi13optimal}. In the model, the FIS is utilized to map the relationship between the input energy efficiency as well as the service success probability and the switching policy. In the QL model, the reward is defined as the weighted probability of a D2D link success probability, while a threshold of cellular link success probability is adopted to decide whether the reward is positive or negative. With the reward function, the $\epsilon$-greedy algorithm allows to explore and exploit the potential switch on/off policies until convergence. Even though every MBS/FBS decides the switching scheme, the control functionality including the initialization and termination of the optimization process is deployed in a central entity. And after each state transition process, MBSs and FBSs receive the overall shared reward determined by the central entity, and uses it to update the Q value to avoid the local selfish optimization.

Lee et al. consider the joint cell activation and user association for load balancing and energy saving in their work~\cite{lee20joint}. The authors adopt the QL method. Specifically, each BS is treated as an agent, while the state and action are current activation variable and mode, respectively. Once each BS chooses an action, a user association scheme can be found by relaxing the load balancing problem to a convex problem. Then, the Q-value based on the heterogeneous network (HetNet) power consumption can be calculated to evaluate the pair of BS activation and user association scheme. By iterating the process until the threshold is reached, the best scheme which jointly optimizes the load balancing and energy efficiency can be obtained. Results illustrate the significant improvement of the network performance and energy efficiency.

\subsubsection{User Association and Load Balancing} To switch the idle BSs to sleep or off mode may result in the overloaded usage of nearby working BSs, which further leads to the QoS deterioration. To strike the balance between energy efficiency and QoS, AI-based user association schemes have been studied.

Zhang et al. adopt the QL technique to decide the user offloading policy to reduce energy consumption as well as improve network throughput~\cite{zhang20dynamic}. In this paper, the authors consider that part of the connected users for each SBS can be offloaded to neighbor SBS or MBS in the multi-tier Ultra Dense Networks (UDNs). In this way, the idle SBS can be turned to sleep or off mode, while the overloaded SBS can be alleviated to ensure the provided services. The proposed QL model aims to solve the problem of how much workload of each SBS can be offloaded to other BSs. The state space includes the load of studied cell and neighbor cells as well as the proportion of users who could be offloaded. And to guarantee energy saving performance and network throughput concurrently, the reward function considers the EE, throughput, and the load difference among the cells. The authors also utilize the mean normalization method to eliminate the sample difference of the considered factors to define the reward function. 

The authors of~\cite{wang19reinforcement} combine the game theory and RL technique to solve the user association and Orthogonal Frequency Division Multiple Access (OFDMA) tile assignment. Specifically, each player is treated as a player to choose the heterogeneous NodeB (hgNB) considering the potential profit and the effects on other players. Since the combinatorial problem can result in the huge size of potential solutions, the authors propose two RL approaches to intelligently guide the search: the regret learning-based algorithm and the fictitious play-based algorithm. In the former one, the Q value is defined according to the regret which is interpreted as the difference between the actual payoff the agent realizes and the potential payoff if another HeNB is chosen. In the latter one, the agent reinforces a strategy considered the payoff calculated on the empirical frequency distribution of the opponents. 

Wang et al.~\cite{wang19machine} utilize the ML techniques to predict the potential traffic burst and then conduct the traffic-aware vehicle association. In their proposal, the supervised learning model is adopted to analyze the statistical correlation between past and present traffic. And online learning is adopted with the goal of minimizing regret instead of loss. In the proposed architecture, every AP performs independent traffic prediction, while the central coordinator conducts the global traffic balance. Since the vehicles are traveling across the APs, the traffic changes in adjacent cells are correlated. Thus, the traffic prediction of each AP is based on the historical data rates and association information of neighboring APs. Once the central coordinator obtains the traffic forecast results, it can proactively update the BS configurations to change the user association information. Thus, some BSs can make preparations for the coming traffic burst, while other BSs can be switched to off mode.

\subsection{Power Control and Resource Allocation}
\label{power-control}
%~\cite{sun18energy} game theory, power control, interference management

\begin{figure*}[!t]
	\centering 
	\subfloat[The transfer learning model for non-stationary wireless channels.]{\label{transfer-learning-a}  \includegraphics[height=8.0cm]{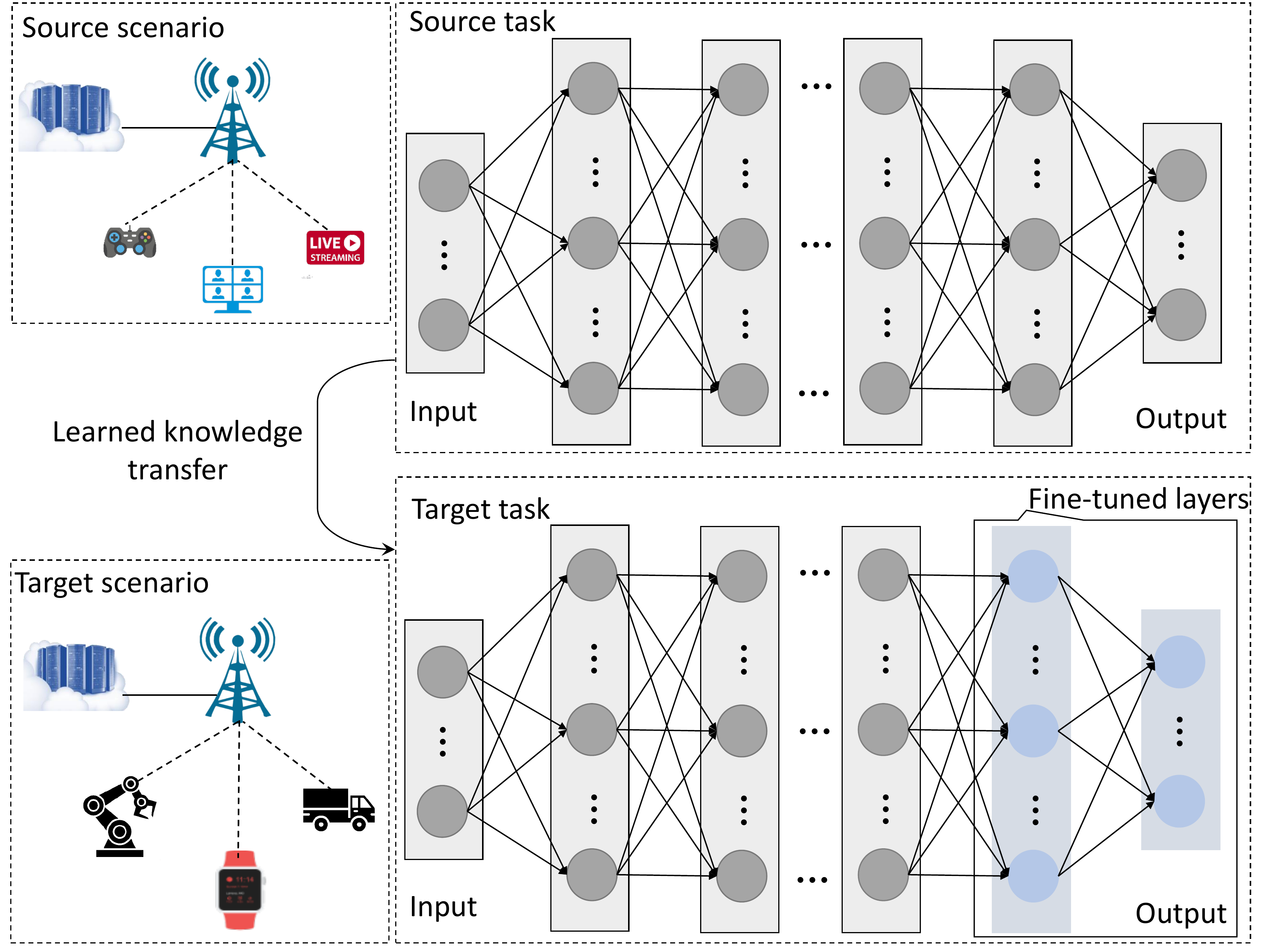}} \hspace{2ex} 
	\subfloat[The transfer learning model for multiple types of services.]{\label{transfer-learning-b}  \includegraphics[height=8.0cm]{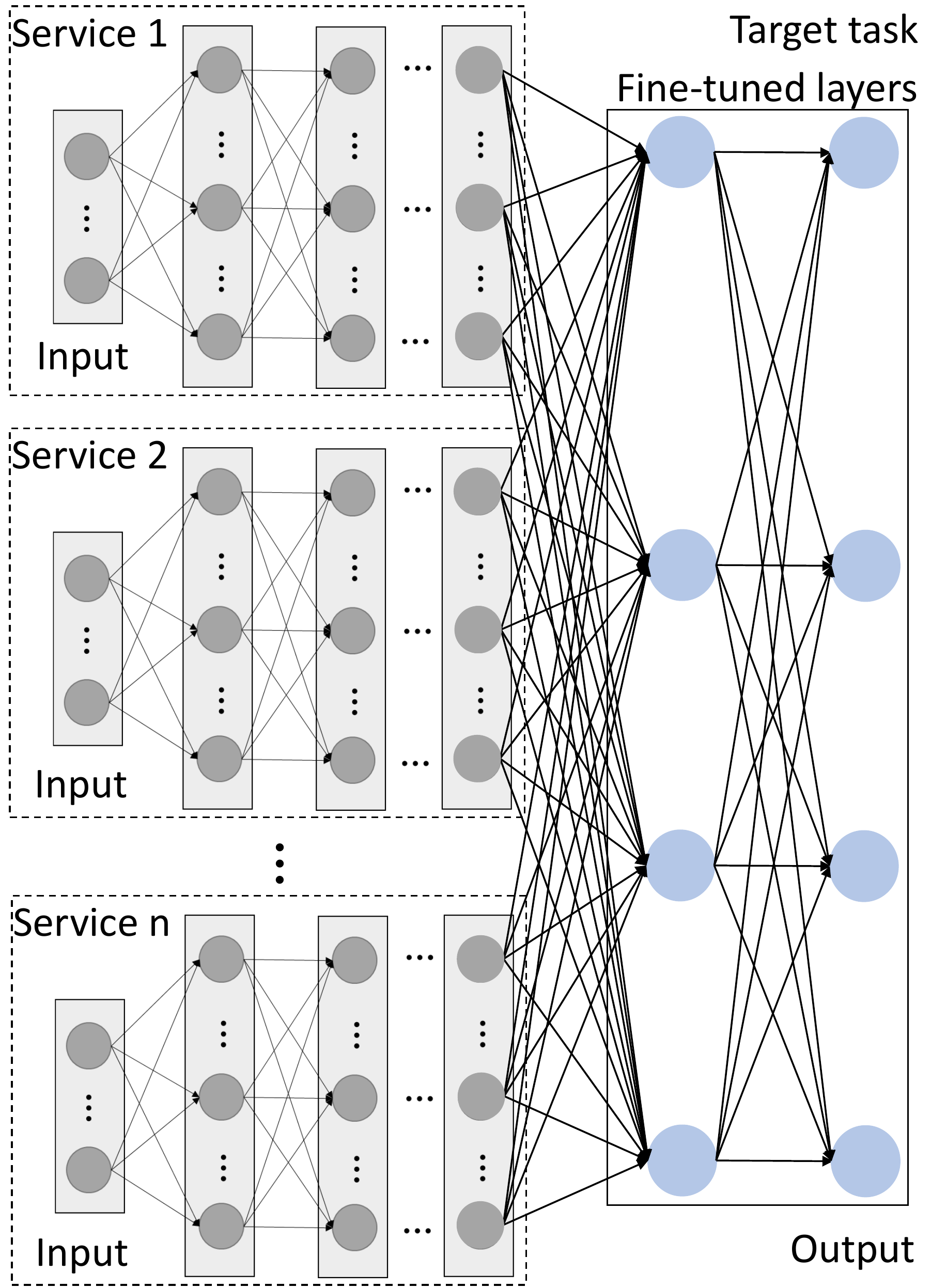}} 
	\caption{The transfer learning techniques for dynamic channel conditions and multiple service types.}
	\label{transfer-learning-model}
\end{figure*}

According to Equation~\ref{energy-efficiency-user}, to improve the system energy efficiency, the transmit power control and resource allocation which affects the interference is critical. Since the ultra massive Multiple-Input Multiple-Output (MIMO), Non-Orthogonal Multiple Access (NOMA), and beamforming technologies will be important techniques in 6G~\cite{khaled19roadmap}, we will introduce the power control for these parts as well as the general power control issue.
\subsubsection{General Power Control}
The transmit power of BSs affects the received SINR at the targeted receivers as well as interference for users in neighboring cells. Thus, the optimization of energy consumption is also jointly considered with interference mitigation through the transmit power control. In~\cite{zhang18reinforcement,xiao20reinforcement}, Zhang et al. utilize the RL technique to optimize the transmit power for alleviating the interference in neighboring cells according to the received SINR and user density. In their proposal, for each transmit power level, every target BS is assumed to obtain a defined utility according to the received SINR at the target users, energy consumption, and interference to non-served users. Then, the Q-value can be defined according to the utility to measure the overall performance of the transmit power level. With the Q-function, the target BSs apply the $\epsilon$-greedy policy to determine the optimal transmit power level. The performance illustrates the reduced energy consumption and interference as well as improvement of network throughput. In~\cite{xiao20reinforcement}, the authors further proposed a CNN based DRL model to map from the network states including the received SINR, user density in the target cell, and estimated channel conditions in neighboring cells, to the transmit power level. The performance illustration shows that the DRL based method can further improve the network performance in terms of energy consumption, throughput, and interference. Another important advantage is that the DRL method converges much faster than the RL based strategy.

Dong et al.~\cite{dong20deep} utilize the Fully-connected NN (FNN) and cascaded NN to optimize the transmit power and channel allocation aiming at minimizing the network energy consumption considering the various service requirements. In this paper, the arrival rates of services and packets are considered as input. For the FNN, the transmit power and channel allocation are adopted as the output. Since the transmit power is a continuous parameter while channel allocation has discrete values, the quantization error in the output layer cannot guarantee the optimal solution even though the DL structure is supervised trained with the labeled data generated by global optimization method. To solve this problem, the authors consider the cascaded FNN structure where the first FNN is to predict the channel allocation and the second for power control of each user. The authors also analyze the non-stationary channel conditions and different service types, and then adopt the transfer learning technique to only fine-tune the last a few layers of the structures through backpropagation process as shown in Fig.~\ref{transfer-learning-model}. For the non-stationary wireless channels, the first FNN in cascaded structure only needs to fine-tune the last a few layers with a small number of data samples as shown in Fig.~\ref{transfer-learning-a}. On the other hand, for the reason that the channel distribution which is the input changes, all layers of the second FNN need to be fine-tuned. Moreover, the authors mention that to fine-tune the last a few layers can be also applied when the service type changes. For instance, the parameters of last a few layers of the cascaded FNN using for delay-tolerant service can be fine-tuned to fit the delay-sensitive or URLLC services as shown in Fig.~\ref{transfer-learning-a}. Furthermore, if we consider multiple types of services exist, the authors propose a structure as shown in Fig.~\ref{transfer-learning-b}, where a few layers are just cascaded at the end of FNN for each service. In this way, we can only fine-tune the parameters of the newly-added layers with a few training samples.

Mattiesen et al.~\cite{matthiesen20global} utilize the ANN to determine the transmit power according to the channel states. The research goal of their proposal is to optimize the weighted sum energy efficiency, which is a non-convex problem. To solve this problem, they first propose an improved Branch-and-Bound (BB) based algorithm to obtain the global optimum solution. Then, the results obtained with this method can be further utilized to train the ANNs in a supervised manner. Since the training is conducted offline, the ANN can be trained with a large dataset generated by the proposed BB-based algorithm to achieve global optimal performance. And the online calculation of the transmit power based on the ANN is illustrated to be robust against mismatches between the training set and real dataset conditions.% a power control

Liu et al.~\cite{liu19power} study the power allocation in a distributed antenna system and utilize the KNN model to optimize the spectrum efficiency and energy efficiency. In this paper, the single-cell distributed antenna system with multiple Remote Access Units (RAUs) is considered and the transmit power of the RAUs should be optimized. However, the research purpose is not for further improvement over traditional methods. On the other hand, they target on solving the high computation overhead of existing methods and hope to utilize the KNN to map the relationship between the user location and power allocation with the assumption of available Channel State Information (CSI) and orthogonal channel resource. Thus, they utilize the traditional method to obtain some data samples for training the KNN models. In the running phase, Euclidean distance between users in the testing and training groups are calculated. And the same power of the nearest neighbor in the training samples is copied to the user in the test group. The final performance analysis shows the KNN can achieve near-optimal performance.

The power control for multi-layer HetNet is more complex and difficult to reach the global optimum. Zhang and Liang~\cite{zhang20deep4} propose a multi-agent-shared-critic DRL method conducted in the core network. Specifically, in the core network, an actor and target actor DNN are trained for every BS, while a shared DNN pair acts as the critic and target critic. The actor DNNs are trained with redundant experience, then share the weight parameters with the corresponding local DNNs. The local DNNs can calculate the transmit power with the real-time local data. To avoid the problem of involving the local optimum, the core network utilizes the global experience to train the critic DNNs. Li et al.~\cite{li18energy2} combine the graph theory and RL technique. In this research, the conflict graph constructed according to the received SINR by the users is utilized to dynamically cluster the cells in order to optimize the channel allocation. To optimize the power control in cell clustering, the RL technique is utilized where the SBS acts as the agent. The state space consists of the interference set and RSS, while the reward is defined according to the throughput and interference.

With the extension of utilized frequency bands to THz, the propagation loss and penetration loss will become increasingly serious. To solve this problem as well as keep satisfied coverage, the radius of future THz-enabled BSs will be limited to 10 meters. Thus, the power control to mitigate the interference in an indoor network will attract increasing attention. Authors in~\cite{gao17qlearning} propose the QL-based distributed and hybrid power control strategies to optimize the network performance in terms of throughput, energy efficiency, and user experience satisfaction. For the BSs without mutual communications, each BS acts as the agent to determine the power for each Resource Block (RB) in a selfish manner. On the other hand, if a central controller is provided, it conducts the QL model to decide the transmit power for each BS. In these two methods, the state is the received SINR level and current transmit power level, while the action is the power level that can be assigned to each RB. The reward functions are defined according to the throughput. 

\subsubsection{Beamforming}
Adaptive beamforming is an important technology to adjust the directionality of the antenna array to enable highly directional transmissions in densely populated areas. Through the adaptive beamforming technique, the network performance of the hotspot can be significantly improved, which further results inincreased energy efficiency. However, the hotspot areas are not fixed due to the dramatically changing user distribution caused by the lifestyle and habits. In~\cite{liu20deep}, Liu et al. utilize the LSTM to extract the spatial and temporal features of UE distributions from the history dataset and detect future hotspots. Based on the location information of predicted hotspots, hybrid beamforming which combines the digital and analog beamforming techniques at the MBS can be adjusted to minimize the total power consumption. Specifically, in the analog beamforming design of massive MIMO systems, the phase shifter can be adjusted to maximize the large array gain. For hybrid beamforming, the optimal power allocation and beamforming directions can be found by converting the original problem into a convex one. The final results also illustrate the reduced energy consumption.

Du et al. jointly optimize the cell sleeping control and beamforming operation by DNN models in~\cite{du19deep}. The authors firstly model the power minimization problem through joint cell sleeping and coordinated beamforming. And the formulated power minimization problem is constrained by the required SINR and maximum power threshold. To alleviate the computation overhead of the numerical method for large-scale scenarios, the authors consider the DNN models to map the relationship between the channel coefficients and beamforming vectors. And the numerical method can be adopted to generate the training data which are further utilized to train the constructed DNN models. To illustrate the performance, the no sleep control and equivalent association strategies are compared. The final results show the DNN-based method can achieve obvious advantages in terms of power saving and satisfactions of QoS demands. 

The authors of~\cite{zhou20manifold} consider the manifold learning~\cite{zheng09manifold} and K-means method~\cite{visalakshi09kmeans} to cluster the multi-cell users into several regions and reduce the complexity of the considered massive MIMO operation. In the two-tier massive MIMO system, the interference mitigation and MIMO hybrid precoding process are challenging due to the large channel dimensionality and high complexity caused by the large antenna count. To alleviate the computation overhead, the authors first utilize the maximum-minimum distance-based K-means method to cluster users into different groups. Thus, with the manifold learning, the nonlinear high dimensional channel coefficients can be transformed into the linear combinations of neighborhood channel coefficients, resulting in significant dimension reduction of the channel matrix while keeping the original geometric properties of the underlying channel manifold. Furthermore, the two-tier beamformers are mainly characterized by the distribution of low-dimensional manifolds and split into outer beamformer and inner beamformer, which are utilized to minimize the inter-cell interference and multi-user intra-cell interference. The final results illustrate the improved SINR and reduced computation complexity. 

Beamforming is also jointly optimized with some other network factors to improve energy efficiency, such as the relay operations. Zou et al.~\cite{zou20optimize,gong20optimize} adopt the DRL technique to improve the multi-antenna Hybrid AP (HAP) beamforming strategies and RF-powered relay operations. In their considered scenario, the individual relay can forward or backscatter the signal to improve the received SINR. Moreover, the relay needs to harvest part of the received power to keep continuous working states. Then, a hierarchical Deep Deterministic Policy Gradient (H-DDPG) model is proposed to select the relay mode and optimize the parameters including the beamforming vector, power splitting ratio, and reflection coefficient in order to maximize the SINR. Specifically, the considered model separates the studied problems into two sub-problems. The DQN model is utilized in the outer loop to select the relay mode. Once the relay mode is selected, the channel conditions, which can be used by the AC networks~\cite{konda00actor} of the inner loop Deep Deterministic Policy Gradient (DDPG) to generate the actions, representing the values of beamforming and relay operation parameters. To accelerate the convergence of the conventional DDPG model caused by the random initialization of double Q-networks, an optimization model is developed to approximate the original problem, which can estimate a lower bound of the target value. The simulations show the improvement of the final reward value and convergence speed compared with the model-free DDPG method. Moreover, the H-DDPG-based framework can significantly improve throughput. 

Since the UAVs are usually adopted as the flying BSs, AI technique has also been applied for the UAV-enabled cellular networks. Li et al.~\cite{li20millimeter} combine the ML and Meaning Field Game (MFG) techniques to jointly optimize the beamforming and beam-steering to maximize the system sum rate. In the considered scenario, to optimize hybrid beamforming lies in the optimization of the hybrid analog beamformer. The Cross-Entropy (CE) function is adopted to evaluate the obtained system sum-rate corresponding to each randomly generated hybrid analog beamformer. In the beam-steering optimization process, the MFG framework is adopted, where the beams act as the agents and information interactions are converted into the interactions with the mass. Considering the conventional numerical methods require a large dimension of action and state spaces, the RL technique is adopted to solve the MFG. Specifically, the state is defined as the combination of index offset of antenna elevation and azimuth angles, while the actions represent the beam selectable path, elevation Angle of Departure (AoD), and aimuth AoD. The reward function is defined according to the obtained system rate. Through the QL process, the optimal action can be chosen. 

\subsubsection{MIMO} In the distributed massive MIMO systems, the pilot sequences transmitted by users are usually adopted to estimate the CSI. However, the pilot contamination caused by the adopted same orthogonal pilot sequences affects the channel estimation accuracy. To alleviate the pilot contamination, the power allocated to each pilot sequence is important. Xu et al. design an unsupervised learning method to predict the power allocation scheme according to the large-scale channel fading coefficients~\cite{xu19deep}. In their research, the authors consider the Minimum Mean-Square Error (MMSE) channel estimator and formulate the problem as the sum MSE minimization. Then, a DNN is exploited with the channel fading coefficients and power allocation as the input and output, respectively. With the loss function defined by the sum MSE of channel estimation, the training process enables the DNN to map nonlinear relationship from channel fading coefficients to the optimal pilot power allocation. Similarly, the authors of~\cite{bashar20exploit} consider the same input and output for the designed Deep Convolutional Neural Network (DCNN). The authors focus on the maximum sum rate problem in limited-fronthaul cell-free massive MIMO. And a heuristic sub-optimal approach is proposed to obtain some data samples, which are to train the DCNN model. Another similar research~\cite{andrea19uplink} is to utilize the ANN to map from the users' positions or shadowing coefficients to the power allocation vector. All of these research works have verified the advantages of DL techniques over traditional mathematical models in terms of the power allocation in massive MIMO systems.

Intelligent power control has also been considered to suppress the attack motivation for more secure communications of MIMO transmitter in~\cite{nie20energy}. In the considered scenario, the malicious attacker can choose different attack modes including jamming, eavesdropping, and spoofing according to the potential reward. The authors combine the game theory and RL to control the power of MIMO transmitter for the suppression of the attack motivation considering the required EE. Specifically, a game model is formulated between the MIMO transmitter and the  malicious attacker. And the RL technique is adopted to derive the optimal power control and transmission probability to reach the Nash Equilibrium (NE) in favor of the MIMO transmitter. The final results illustrate the improvement of transmission secrecy performance and energy efficiency.

The authors of~\cite{gao17machine} and~\cite{dong18energy} utilize the CE-based algorithm to solve the hybrid procoding problem in mmWave massive MIMO systems. Specifically, the CE-based algorithm is adopted to update the probability distribution of the analog beamformer in the iteration process and then the "elite" analog beamformer which can result in minimum total transmit power can be found. Moreover, the authors of~\cite{dong18energy} adaptively weight different elites according to their objective values, which can further improve the performance of CE-based algorithms. The simulations in the two paper verify the CE-based hybrid precoding scheme can improve energy efficiency of mmWave massive MIMO systems with low complexity.

Different from the above research focusing the intelligent power control in the massive MIMO system, the authors of~\cite{ge18energy} propose a DL-based user-aware antenna allocation strategy. In their research, the LSTM model trained with the real dataset is adopted to predict the variations of future associated users for the massive MIMO-enabled BSs, which is similar to the applications of DL in traffic forecast~\cite{vallero19green,gao20machine,donevski19neural}. Based on the prediction results, the optimum number of BS antennas are allocated to maximize the EE. 

\subsubsection{NOMA}

The NOMA technique introduces an extra power domain to enable multiple users to be multiplexed on the same channel resource~\cite{he19joint}, which can improve the network capacity and resource efficiency. Thus, the resources including the power and channels are usually considered as the key metrics to be optimized for network performance improvement. In~\cite{he19joint}, the authors first utilize the DRL technique to conduct the channel assignment for alleviating the computation overhead of conventional methods due to the huge solution space. In the proposal, each BS acts as the agent, while the NOMA system is regarded as the environment. The attention-based NN is adopted to model the channel assignment policy, with the encoder computing the embedding of state space and decoder outputting the probability distribution over all states. Once a channel assignment solution is obtained, the corresponding power allocation can be calculated. Then, the derived system performance is further utilized to define the reward function. And the training process enables the proposed NN to find the optimal channel assignment according to the system states with low complexity. The authors of~\cite{yang20deep} also utilize the DL technique to alleviate the computation overhead of conventional methods. However, their proposal train the DNN in a supervised manner, where the downlink channel gains and corresponding power allocation scheme are as the input and output, respectively. 

Zhang et al.~\cite{zhang20deep2} also consider the DL-based radio resource management to improve EE in NOMA networks. Besides the subchannel and power allocation considered in~\cite{he19joint}, the authors of~\cite{zhang20deep2} also analyze the user association since they consider the two-tier networks including MBSs and SBSs. The authors optimize these three factors separately with three methods. Specifically, the semi-supervised learning-based NNs and supervised learning-based DNN are adopted to optimize subchannel assignment and power allocation, respectively, while the Lagrange dual decomposition method is used to solve the user association problem. In the optimization of subchannel assignment, the numerical iterative method, the two-side matching algorithm, is utilized to generate some labeled data samples, which cooperate with the unlabeled data to train the NN in a semi-supervised learning manner. The authors consider the co-training semi-supervised learning model~\cite{blum98combine}, where two NNs are trained with the data from different views to produce the optimal learner. The input and output of the NNs are the channel gains and allocation strategies, respectively. Since the classification with unlabeled data still depends on the labeled data, the authors select the highly confident labeled data with the most consistency. To optimize the power allocation, the DNN is trained with the labels generated by an iterative gradient algorithm. 

User clustering is an important factor to improve energy efficiency for NOMA-enabled multi-tier cellular networks. Zhang et al.~\cite{zhang20energy, zhang20energy2} adopt the K-means clustering to cluster the users in Thz MIMO-NOMA systems. In their research, the users are separated into different clusters of SBSs and MBS in the coverage. Since the THz transmission is challenged by the severe path spreading loss and molecular absorption loss, a suitable clustering scheme can improve the channel quality and suppress the interference, resulting in higher SINR and transmission throughput. Then, the authors propose an enhanced K-means strategy to cluster the users. To overcome the fluctuation with different initial clustering centers for the conventional K-means method, the authors calculate the channel correlation parameters of different cluster heads and choose the one that maximizes the metric. And the MSE analysis clearly verifies the improved convergence compared with the conventional K-means method. 

\subsection{Energy Harvesting-enabled Base Station}
Motivated by the concern for climate change and inspired by the development of energy harvesting, the renewable energy resources have been considered to alleviate the requirement for the power grid. On the other hand, the dynamics of renewable energy resources complicate the management and operation of cellular networks. AI techniques have been widely studied to track the dynamic harvesting source and optimize network operations.

To optimize cellular network performance with renewable energy-enabled BSs, the most direct method is to predict the harvesting power. For the scenario where BS is powered by a photovoltaic (PV) panel, battery, and power grid, the authors of~\cite{vallero19green} adopt the Block Linear Regression (BLR)~\cite{pan15block}, ANN~\cite{lee92short}, and LSTM~\cite{lstm-introduction} to forecast the traffic, while the linear regression model is utilized to predict the dynamic harvesting power. To measure the performance of these ML models, the metrics including the Average Mean Absolute Relative Error (AMARE) and Average Mean Error (AME) are analyzed. Then, the prediction results can be utilized to switch off some micro BS in low usage to save energy. 

Miozzo et al.~\cite{miozzo17switch} propose a distributed RL-based SBS switching strategy to balance the network drop rate and energy consumption for two-tier cellular networks where the SBS and MBS are powered by the electricity grid and renewable solar energy. The state space includes the instantaneous energy harvested, battery level, and traffic load, while the reward is defined according to the system drop rate and battery level. However, this method has the limitation to reach the system optimization since each SBS acts as the agent and decides the working state according to its local state. To alleviate this problem, the authors further propose a layered learning optimization framework in~\cite{miozzo20coordinate}. In the lower layer, each SBS still follows the original manner to decide the switching scheme in a distributed intelligent manner. The only difference is that a heuristic function is defined and united with the regular Q-value to select the optimal policy. Moreover, the heuristic value is decided in the upper layer in a centralized manner. Specifically, the MBS utilizes a multi-layer NN to forecast its traffic load and judge whether the system is under-dimensioned or over-dimensioned. Based on the load estimation, the heuristic value is derived. 

Li et al.~\cite{li18deep3} utilize the DRL method to manage the work states of the harvesting-enabled SBS in a centralized manner. In their proposal, the central controller acts as the agent to decide the action which is  a vector consisting of binary units representing the switching decision for each SBS. And the state space includes the harvested energy, battery levels, traffic loads, throughput, and delay of all SBSs. Since the research aims to balance the EE and QoS, the reward function is defined as the weighted sum of the two metrics. Using the DNNs to approximate the Q-value, the final simulation results clearly illustrate the advantages of DQL against the traditional QL in terms of energy efficiency and delay. On the other hand, this method has a shortcoming that the size of action space exponentially increases with the number of SBSs, which leads to abundant explorations during the training process. To solve this problem, Li et al.~\cite{li18deep2} consider the DDPG model. In this model, the AC algorithm~\cite{actor-sergey} is adopted where an actor NN and a critic NN are adopted to select an action and evaluate the selected action, respectively. The final results verify the improved energy efficiency over DQN and QL methods.

Since the renewable energy-enabled BSs are usually equipped with batteries to store the harvested energy, to optimize battery management can also contribute to the EE. The authors of~\cite{mendil18battery} propose the FQL-based power management which combines the QL and FIS~\cite{busoniu07fuzzy} to minimize the electricity expenditures and enhance the battery life span. The authors also construct the power consumption model related to the real-time traffic as well as the battery aging model, which is meaningful to design a more detailed energy-efficient BS management policy in the future. Piovesan et al.~\cite{piovesan20joint} analyze the constrained capacity of SBS battery and consider energy sharing in the design of the SBS switching scheme. The authors utilize and compare imitation learning~\cite{imitation-learning-tutorial}, QL, and DQL methods. The considered state includes the battery level and harvested energy, while the reward functions in two RL models are defined according to the grid energy consumption. In the imitation learning model, the ANN is supervised trained with the labeled data generated by a mathematical model~\cite{piovesan18optimal} to map the relationship between the system state and switch action. For the two RL models, the difference is that the Q-value is stored in a table for QL, while DQL utilizes an ANN approximator to estimate the Q-value. The final comparison illustrates the DQL model achieves the best performance in terms of energy saving and system outage, which is more suitable for the highly-dense scenarios.

Wei et al.~\cite{wei18user} utilize the policy gradient-based AC networks~\cite{grondman12efficient} to solve the user scheduling and resource allocation problem for the optimization of EE in a two-tier HetNet where the SBSs are powered by solar and wind energy. Since the wireless fading channels and stochastically harvested renewable energy have the Markovian property~\cite{simsek15learning}, the optimization of user scheduling and resource allocation can be formulated as an MDP, which lays the foundation for using DRL method. In their proposal, the state space consists of the SINR of each user and battery energy level of each SBS, which are both continuous variables. The action space includes the number of allocated users and subchannels as well as the transmission power. The reward function is defined as the EE with only the grid energy consumption considered. And through online training, the final numerical analysis illustrates the improved EE.

From the above introductions, it can be found that AI techniques are efficient to address the dynamics of energy harvesting process. And similar to the BSs which are only powered by electricity grid, AI models can be utilized to optimize the switching scheme, user association, power control, and resource allocation. 

\subsection{summary}
In the above research, AI techniques can be utilized to optimize different network parameters in order to reduce energy consumption or improve the EE. The supervised learning technique can be utilized to regress the complex unknown relationships among the network parameters. For example, AI models can be trained with the data generated by conventional methods to map the relationship between channel conditions and power allocation~\cite{du19deep,bashar20exploit}. Thus, AI-based algorithms can avoid the massive iterations and alleviate the computation overhead of conventional methods. Moreover, the RL and DRL techniques can efficiently address the problem of the huge size of solution space~\cite{zhang20deep,li18deep2}. Furthermore, the combinations of ML/DL models with heuristic algorithms or game theory can further enhance efficiency~\cite{dai19hybrid,ho13online,moysen16machine,wang19reinforcement,nie20energy}

\section{Machine Type Communications}
\label{mtc-section}
Besides the cellular networks, MTC techniques provide users with more choices and flexibility. And the development of IoT will result in a great surging number of MTC devices~\cite{kawamoto19multilayer}. In this section, we first give the power consumption model of MTC and introduce the related AI strategies to reduce energy consumption and improve efficiency.

\subsection{Power Consumption Modeling}

The actual energy consumption of MTC depends on the definite scenario including the transmission policy, devices, information size, and so on. In this part, we give a general power consumption model for the single-hop MTC scenario, by which the multi-hop power consumption model can be derived.

The total power consumption of a machine node is mainly utilized for two purposes: transmission and receiving packets, which can be simplified in the following equation. The details can be referred to~\cite{wang06realistic}.
\begin{equation}
\label{power-consumption-sensor}
P_{m}(d)=P_{t0}+P_{r0}+P_a(d)
\end{equation} 
\noindent{where $P_{m}$ denotes the total power consumption of an MTC node. $P_{t0}$ and $P_{r0}$ are the power consumed by the circuit for transmitting and receiving and usually regarded as constants. $P_a(d)$ denotes the power consumption of Power Amplifier (PA), where $d$ is the transmission distance. From the equation, it can be found that the total power consumption depends on the PA. However, the value of $P_a(d)$ is affected by many factors including the specific hardware implementation, DC bias condition, load characteristics, operating frequency, and the required PA output power $P_{mt}$. In a specific scenario with given MTC devices, we usually only study the required minimum PA output power while the other factors are constant. And the relationship between the two metrics can be denoted as below:}
\begin{equation}
\label{transmission-power}
P_{mt}(d)=\eta{P_a(d)}
\end{equation}
\noindent{where $\eta$ denotes the drain efficiency of PA. Specifically, the value of required minimum PA output power $P_{mt}$ can be calculated according to the given SINR threshold at the receiver side and the path loss model between the transmitter and receiver. Then, the power consumption for a single-hop MTC model can be calculated.} By adding the power consumption for each hop, the multi-hop power consumption can be obtained. Since the definition of energy efficiency in MTC is similar to that in cellular networks, we can use Equation~\ref{energy-efficiency-user} accordingly. 

It can be found the power consumed by the MTC node is mainly to support the circuit and PA. Since most of the MTC nodes do not need to keep the working state, the idle nodes can be turned into the sleep state to reduce the circuit energy consumption. For the working nodes, how to reduce the required transmit power $P_{mt}$ as well as minimize the transmission time are the main factors considered in green communications. For the former part, the transmit power depends on the path loss and required SINR at the receiver. The practical solutions to reduce the transmit power include the optimization of network deployment, access technologies, and resource allocation. To reduce the transmission time, we need to optimize the transmission protocols, such as routing and relay. Similar to the renewable energy-enabled BSs, energy harvesting and sharing are also important techniques toward green MTC. The following paragraphs introduce the related research one by one.

\subsection{Energy-Efficient Network Access}
Various access technologies have been developed for different MTC scenarios, such as cellular communications, IEEE 802.15.4, WiFi, Narrow-Band IoT (NB-IoT), backscatter communications, and so on. The satellites and Unmanned Aerial Vehicles (UAVs) have been emerging platforms to provide Internet access for devices. In this part, we discuss how AI is utilized to improve energy efficiency of these access technologies. 
\subsubsection{Terrestrial Access Configurations} Even though cellular communications can provide stable and high-throughput connections, the high power consumption to keep connections as well as the expense of the cellular infrastructure challenge the wide applications in MTC. Moreover, since different MTC services have heterogeneous QoS requirements and are distributed in various areas including the sparsely populated areas and hazardous environments, to develop corresponding  access techniques is important to reduce energy consumption or extend the lifetime. 
Some AI researcher works related to improve energy efficiency of these access technologies are introduced in the following paragraphs. We also give Table~\ref{ai-access} to give more examples to adopt AI to optimize the access layer for green communications.

Li et al. adopt the RL technique to optimize the duty cycle control for each router node in IEEE 802.15.4-based M2M communications~\cite{li15qos,li15smart}. The authors consider the QL method to design the superframe order for minimizing the sum of weighted energy consumption and delay. In the considered RL model, the agent interacts with the environment and chooses the suitable superframe order according to the queue length. And the final simulation results verify the improved energy efficiency. Xu et al. also utilize the model-free RL method to improve the throughput and EE of IEEE 802.15.4-enabled Industrial IoT (IIoT) networks~\cite{xu20priority}. In their research, the QL is adopted to adjust the sampling rate of the control subsystem and backoff exponential, which is difficult to be addressed by traditional stochastic modeling approaches. For the IEEE 802.15.4-based MTC scenarios, Zarwar et al.~\cite{sarwar20reinforcement} give a comprehensive survey on RL-enabled adaptive duty cycling strategies, which can be referred for more knowledge.

Alenezi et al. focuses on LoRa communication technology and utilize the K-clustering method to cluster the nodes in order to reduce the collision rate~\cite{alenezi20unsupervise}. To address the high probability of packet collision caused by random access and simultaneous transmissions, the authors first utilize the K-means clustering to separate the IoT nodes into several groups and then schedule their transmissions according to dynamic priority. The final simulations illustrate the significant reduction of collision rate, which further results in the decreased transmission delay and energy consumption. Azari and Cavdar~\cite{azari18organize} also utilize AI to optimize the performance of LoRa. The authors consider the Multi-Agent Multi-Arm Bandit (MAB) to choose the best transmit power level, spreading factor, and subchannel to maximize the reward which is defined as a weighted sum of communication reliability and EE. The analysis illustrates the lightweight complexity of the proposed algorithms and verifies the performance improvement in terms of energy efficiency and transmission success probability.

\begin{table*}[]
	\begin{tabular}{|c|c|c|c|c|c|c|}
		\hline
		Research work                                                                 & \begin{tabular}[c]{@{}c@{}}Access\\ technology\end{tabular}      & \begin{tabular}[c]{@{}c@{}}Learning\\ method\end{tabular} & \begin{tabular}[c]{@{}c@{}}AI\\ model\end{tabular} & \begin{tabular}[c]{@{}c@{}}Input/\\ state\end{tabular}                                         & \begin{tabular}[c]{@{}c@{}}Output/\\ action\end{tabular}                & \begin{tabular}[c]{@{}c@{}}Target/\\ metric\end{tabular}                \\ \hline
		\begin{tabular}[c]{@{}c@{}}Zhou et al.\\~\cite{zhou20deep}\end{tabular}      & UAV                                                              & reinforcement                                             & DNN                                                & \begin{tabular}[c]{@{}c@{}}UAV location,\\ task information,\\ energy cost\end{tabular} & \begin{tabular}[c]{@{}c@{}}offloading\\ decision\end{tabular}           & \begin{tabular}[c]{@{}c@{}} delay\end{tabular}         \\ \hline
		\begin{tabular}[c]{@{}c@{}}Nguyen et al.\\~\cite{nguyen20real}\end{tabular}  & UAV, D2D                                                          & reinforcement                                             & DDPG                                               & \begin{tabular}[c]{@{}c@{}}D2D SINR \\ information\end{tabular}                                & \begin{tabular}[c]{@{}c@{}}harvesting\\ time\end{tabular}               & \begin{tabular}[c]{@{}c@{}}energy \\ efficiency\end{tabular}            \\ \hline
		\begin{tabular}[c]{@{}c@{}}Chen et al.\\~\cite{chen20optimum}\end{tabular}   & satellite                                                        & reinforcement                                             & QL                                                 & \begin{tabular}[c]{@{}c@{}}traffic demand,\\ channel condition\end{tabular}                    & \begin{tabular}[c]{@{}c@{}}power\\ allocation\end{tabular}              & \begin{tabular}[c]{@{}c@{}}service\\ fairness\end{tabular} \\ \hline
		\begin{tabular}[c]{@{}c@{}}Özbek et al.\\~\cite{ozbek20energy}\end{tabular}  & \begin{tabular}[c]{@{}c@{}}cellular \\ network, D2D\end{tabular} & supervised                                                & ANN                                                & channel gain                                                                                   & \begin{tabular}[c]{@{}c@{}}power\\ allocation\end{tabular}              & \begin{tabular}[c]{@{}c@{}}energy\\efficiency\end{tabular}       \\ \hline
		\begin{tabular}[c]{@{}c@{}}Zhang et al.\\~\cite{zhang20energy3}\end{tabular} & \begin{tabular}[c]{@{}c@{}}cellular \\ network, D2D\end{tabular} & reinforcement                                             & DDPG                                               & \begin{tabular}[c]{@{}c@{}}QoS satisfaction\\ degree\end{tabular}                              & \begin{tabular}[c]{@{}c@{}}mode selection,\\ power control\end{tabular} & \begin{tabular}[c]{@{}c@{}}energy\\efficiency\end{tabular}  \\ \hline
		\begin{tabular}[c]{@{}c@{}}Ji et al.\\~\cite{ji20power}\end{tabular}         & \begin{tabular}[c]{@{}c@{}}cellular \\ network, D2D\end{tabular} & reinforcement                                             & DQN                                                & \begin{tabular}[c]{@{}c@{}}SINR\\ transmission power\end{tabular}                              & \begin{tabular}[c]{@{}c@{}}transmission\\power change\end{tabular}     & \begin{tabular}[c]{@{}c@{}}energy\\efficiency\end{tabular}  \\ \hline
		\begin{tabular}[c]{@{}c@{}}Chowdhury et al.\\ ~\cite{chowdhury19drift}\end{tabular} & unknown                                                          & reinforcement                                             & DNN                                                & \begin{tabular}[c]{@{}c@{}}current resource\\ allocation, service\\ deman\end{tabular}         & \begin{tabular}[c]{@{}c@{}}resource \\ allocation\\ strategy\end{tabular} & \begin{tabular}[c]{@{}c@{}}energy cost\\ and delay\end{tabular} \\ \hline
		\begin{tabular}[c]{@{}c@{}}Yang et al.\\~\cite{yang20learn}\end{tabular}                           & WiFi, VLC                                                        & reinforcement                                                        & DQN                                                & \begin{tabular}[c]{@{}c@{}}channel status,\\ channel quality\\ service types,\\ service satisfication\end{tabular} & \begin{tabular}[c]{@{}c@{}}network selection,\\ channel assignment,\\ power management\end{tabular}  & \begin{tabular}[c]{@{}c@{}}energy\\ efficiency\end{tabular}                                    \\ \hline
		\begin{tabular}[c]{@{}c@{}}Yang et al.\\~\cite{yang20actor}\end{tabular}                           & CRN                                                              & \begin{tabular}[c]{@{}c@{}}actor-critic\\ reinforcement\end{tabular} & AC                                                 & \begin{tabular}[c]{@{}c@{}}channel states,\\ device priority,\\ channel SNR,\\ traffic load\end{tabular}           & \begin{tabular}[c]{@{}c@{}}power control,\\ spectrum management,\\ modulation selection\end{tabular} & \begin{tabular}[c]{@{}c@{}}transmission\\ rate, power \\ throughput,\\and delay\end{tabular} \\ \hline
		\begin{tabular}[c]{@{}c@{}}Rahman et al.\\~\cite{rahman20learn}\end{tabular}                       & RAN                                                              & reinforcement                                                        & DQN                                                & \begin{tabular}[c]{@{}c@{}}content states,\\ channel states,\\ and power\end{tabular}                              & \begin{tabular}[c]{@{}c@{}}caching policy,\\ and power \\ allocation\end{tabular}                    & delay                                                                                          \\ \hline
		\begin{tabular}[c]{@{}c@{}}Sharma et al.\\~\cite{sharma19distribute,sharma19multiple}\end{tabular} & unknown                                                          & reinforcement                                                        & DNN                                                & \begin{tabular}[c]{@{}c@{}}battery states, and\\ channel gains\end{tabular}                                        & transmit power                                                                                       & throughput                                                                                     \\ \hline
		\begin{tabular}[c]{@{}c@{}}Bao et al.\\~\cite{bao19stochastic}\end{tabular}                        & cellular                                                         & reinforcement                                                        & QL                                                 & battery states                                                                                                     & transmit power                                                                                       & SNR                                                                                            \\ \hline
	
	\end{tabular}
	\caption{Some Related Research Works Using AI to Optimize Network Access for Green MTCs}
	\label{ai-access}
\end{table*}
 
Guo and Xiang~\cite{guo19multiple} utilize the distribute multi-agent RL technique to pick the power ramping factor and preamble for each UE in the NB-IoT networks. In their research, an adaptive learning rate based QL algorithm is proposed for the non-stationary environment, with the reward defined according to the UE's energy consumption. Moreover, the learning rate is adjusted after comparing current expected reward and expectations. The authors of~\cite{jiang19reinforcement} also utilize the QL techniques to optimize the configurations in the random access process. Their proposal focuses on the optimization of three parameters including the number of random access channel periods, the repetition value, and the number of preambles in each access period. In the single-cell scenario, the tabular QL, linear approximation-based QL, and DQN methods are adopted by the eNB to predict the number of preambles in order to maximize the served IoT devices. In the multi-cell scenario, the huge size of the action space composed of three parameters is a great challenge. The authors consider an action aggregated approach by converting the selection of definite value to the choice of increase or decrease. Then, the three QL methods are compared with a cooperative multi-agent DQL proposed. 

Lien et al. study the intelligent radio access in vehicular network to strike the balance among energy efficiency, latency, and reliability~\cite{lien19low}. The authors concentrate on the fronthaul radio resource starvation and propose an RL-based MAB algorithm to avoid the backhaul transmission in the core networks. In the considered scenario, each vehicle can simultaneously access multiple BSs to request the contents using the feedbackless transmission schemes, which further means different communication reliability, energy consumption, and latency. To strike the tradeoff among energy efficiency, latency, and reliability, the authors first formulate the Lyapunov function~\cite{cui12survey} to derive the optimum number of BSs to meet the content request of each vehicle. Then, to decide whether to use the feedback-based or feedbackless transmission scheme, the authors construct the MAB model and utilize the $\epsilon$-greedy RL algorithm to solve this problem. Specifically, the research goal of this step is to minimize the long-term expected cost which is defined as the weighted sum of  request drop event, transmission latency, and energy consumption.

\subsubsection{Access through Satellites} Satellites can provide seamless coverage for IoT devices, especially for rural and remote areas. However, the large path loss challenges the system EE and lifetime. Authors of~\cite{zhao20deep} study DRL-based channel allocation to improve the system EE as well as guarantee the QoS for LEO satellite IoTs. The authors formulate the channel resource allocation as an MDP and further utilize the DRL technique to solve it. In their proposal, the agent is assumed to choose an action to assign the channels according to the state which is defined as the user task size and location. The authors also construct the users' requests into an image as the input of the considered NN, which can reduce the input size and accelerate the learning process. The proposed intelligent approach is illustrated to save more than 60\% of energy.  

Sun et al. utilize the DL technique to optimize the Successive Interference Cancellation (SIC) decoding order for NOMA downlink system in satellite IoT networks~\cite{sun19deep}. In this research, the long-term joint power allocation and rate control scheme is formulated to improve the NOMA downlink rate. Then, the Lyapunou optimization framework~\cite{cui12survey} is adopted to convert the original problem to a series of online optimization sub-problems, where the power allocation depends on the SIC decoding order, which is further affected by the queue state and channel states. Due to the continuous changes, the DNN model is adopted to map from the states of queues and channels to the SIC decoding order. Moreover, the DNN is trained in a supervised manner with the data obtained by traversing all possible choices. 

Han et al. combine the game theory and DRL to optimize the anti-jamming performance of satellite-enabled army IoT networks~\cite{han20dynamic}. In their considered scenario, the sensing devices are separated into different groups and the messages are relayed by the sink nodes to reduce energy consumption. Assuming the smart jammers may launch jamming attacks to the IoT devices, the authors first utilize the DRL technique to select the optimal location of jammers for the maximum jamming effect. In the DRL model, the reward is defined with the estimated transmission energy consumption and minimal value without jamming attacks into account. With the defined jamming policy, the anti-jamming part is constructed as a hierarchical anti-jamming stackelberg game, which is not the focus of this paper. 

\begin{figure}[!t]
	\includegraphics[scale=0.4]{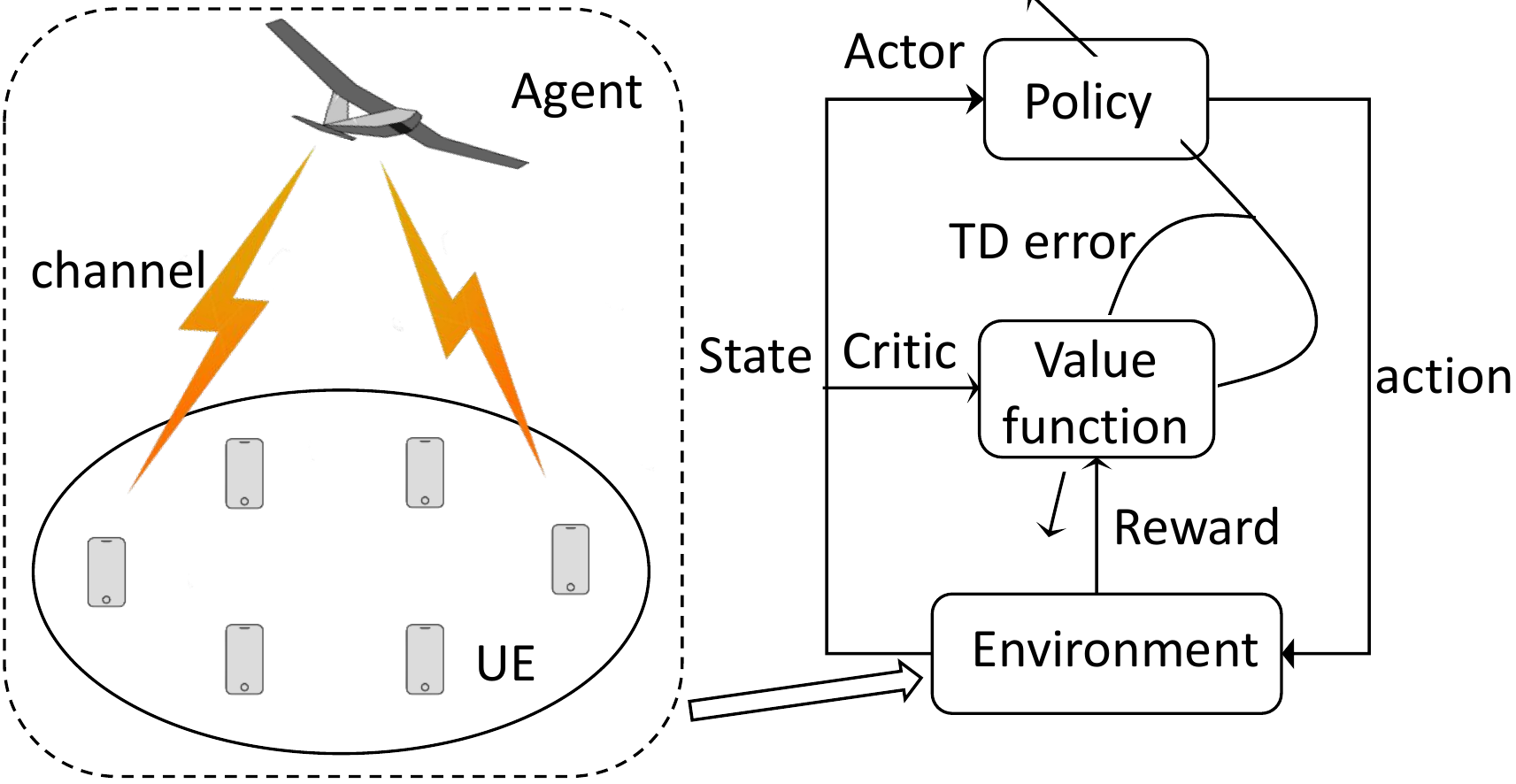}
	\caption{The Actor-Critic model for UAV-assisted IoT network.}
	\label{ac-model}
\end{figure}

\subsubsection{Access through UAVs} Since UAVs can be easily controlled to fly over the communication terminals, they have been widely recognized as air BSs or gateway to provide periodical Internet connections~\cite{liu18energy,khairy20constrain,zhou20deep,cao19deep}. Liu et al.~\cite{liu18energy} consider the UAVs to collect the sensing information aggregated by the collection point from terrestrial IoT end terminals. In the considered scenario, the flying trajectory of UAVs affects the received SINR, which further impacts the number of uploaded packets for each collection point in a round. Then, the authors utilize the DRL technique to optimize the trajectory with the defined reward considering EE and packet delay priority. The dueling DQN is utilized to decide the moving direction according to current states which consist of both the delay priority and energy consumption priority. The simulation illustrates the improved average reward with different network scale and data density.

Cao et al.~\cite{cao19deep} utilize the DRL technique to optimize the channel allocation and transmit power control for the IoT nodes. Specifically, with the fixed trajectory assumed, the UAV acts as the agent to select a suitable channel and transmit power for every IoT uplink at each time slot in order to maximize the reward which is defined as the EE of all IoT nodes. The AC network~\cite{grondman12efficient} is adopted in the DRL algorithm, where the actor and critic NN have different structures as shown in Fig.~\ref{ac-model}. Moreover, the authors also try the different number of trajectory steps to update the channel and power allocations through the simulations. Similar research in~\cite{yuan20energy} also considers the UAV-enabled BS with the predefined trajectory. However, the research aims to optimize user scheduling and hovering time assignment for improving the EE of battery-constrained UAVs. Since the problem is a discrete constrained combinatorial problem that is beyond the conventional AC model, the authors consider the stochastic policy to address the issue of huge discrete spaces. A flexible reward function is defined with an adjustable parameter. The final performance illustrates that the proposed model can save nearly 30\% of energy compared with the conventional AC model. 

The above paragraphs give some typical research works on AI-based network access toward green MTCs. We further list some research works and the utilized AI techniques as shown in Table~\ref{ai-access}.

\subsection{Energy-Efficient Transmissions}
In many MTC scenarios, the messages are transmitted in a cooperative manner from the senders to receivers or APs due to the resource constraints. Then, the routing path design or the relay selection affects both the network performance and power consumption~\cite{liu19reinforcement,wang20optimal}. Different from the message transmission in wired networks, the path and relay selection of MTC scenarios needs to consider  more issues, such as energy dynamics~\cite{wang20multi-hop,wang20optimal}, node mobility~\cite{lien19low,zhang20cooperative}, spectrum efficiency~\cite{liu19reinforcement,zhou18energy}, QoS~\cite{mostafaei19energy,wang20energy}, and even the information security~\cite{haseeb20secure,xiao20reinforcement2}. The following paragraphs and Table~\ref{ai-transmission} show the related research.

\subsubsection{Routing}

Liu et al. study the routing problem in the wide area mesh IoT networks and consider the RL technique to address the limitations of conventional methods in terms of energy sensitivity~\cite{liu19reinforcement}. In their proposal, the model-free RL method called temporal difference learning is adopted to populate and update the routing table. Specifically, the routing metric which indicates the probability of selecting a particular adjacent node is calculated by using a Boltzmann exploration process. And once the routing metric values of the visited nodes in all paths are calculated, the path quality value is computed using the RL method. To improve energy sensitivity of the routing method, the cost function is defined according to the transmission power as well as the remaining energy of transmit and receive nodes. The final simulation results illustrate the performance improvement in terms of energy efficiency as well as the success rate and spectrum efficiency.

The routing design in underwater sensor networks (UWSN) is a hot application of AI techniques. Zhou et al. utilize the QL method to select the next node and define the reward function according to the residual energy and depth information for a balance of End-to-End (E2E) delay and energy consumption~\cite{zhou19learning}. The utilization of QL enables the long-term reward taken into account, which finally reaches the global optimization. By sorting the neighbors according to the calculated Q-value, the node with higher priority can be selected to forward packets, while the other neighbors with smaller Q-values are suppressed for energy saving. Hu and Fei also adopt the QL to solve the routing in UWSNs~\cite{hu10machine}, while the research goal is to make the residual energy of sensor nodes more evenly distributed for the maximum network lifetime. In the RL proposal, the authors consider not only the residual energy but also energy distribution in a group of sensor nodes to define the cost function, which is further utilized to calculate the reward and Q-value for different actions indicating various next nodes. The authors in this paper also illustrate that the proposed method can converge for dynamic scenarios. And final performance results indicate the lifetime can be extended up to 20\%.

In~\cite{aboubakar19toward}, authors adopt supervised learning-based MLP algorithm to improve the routing performance and energy efficiency for the IoT low power networks. Different from the other works~\cite{liu19reinforcement,zhou19learning,hu10machine} which utilize AI models to predict the next node directly,~\cite{aboubakar19toward} aims at optimizing the value of transmission range of each node to improve the routing performance and minimizing energy consumption. In this paper, the authors first construct an IoT network to collect the labeled data including node positions and corresponding transmission range. Then, the MLP is trained with the labeled data to map the relationship from the node position to the optimal transmission range. One of the advantages of this proposal is to address the high dynamics of IoT networks. And the final simulations illustrate the extension of the network lifetime. 

Mostafaei studies the multi-constrained routing problem in WSNs and proposes a distributed learning approach~\cite{mostafaei19energy}, where each node is regarded as a learning automaton. After the initial phase each learning automaton senses the neighbor nodes to construct the action space, it transmits a packet by a randomly selected action. Once the packet reaches the sink node, the environment will feedback a reinforcement signal which can be a penalty or a reward to evaluate the selected action. Then, the transmission probability of each action for every node can be updated.

\begin{table*}[]
	\begin{tabular}{|c|c|c|c|c|c|c|}
		\hline
		\multicolumn{1}{|c|}{\begin{tabular}[c]{@{}c@{}}Research\\ work\end{tabular}}   & \multicolumn{1}{c|}{\begin{tabular}[c]{@{}c@{}}Network\\ scenario\end{tabular}} & \multicolumn{1}{c|}{\begin{tabular}[c]{@{}c@{}}Learning\\ method\end{tabular}} & \multicolumn{1}{c|}{\begin{tabular}[c]{@{}c@{}}AI\\ model\end{tabular}} & \multicolumn{1}{c|}{\begin{tabular}[c]{@{}c@{}}Input\\ /state\end{tabular}}              & \multicolumn{1}{c|}{\begin{tabular}[c]{@{}c@{}}Output/\\ action\end{tabular}} & \multicolumn{1}{c|}{\begin{tabular}[c]{@{}c@{}}Target/\\ reward\end{tabular}}                                \\ \hline
		\begin{tabular}[c]{@{}c@{}}Fu et al.\\~\cite{fu20deep}\end{tabular}            & \begin{tabular}[c]{@{}c@{}}vehicular\\ energy \\ network\end{tabular}           & supervised                                                                     & LSTM                                                                    & traffic flow                                                                             & \begin{tabular}[c]{@{}c@{}}future\\ traffic\end{tabular}                      & \begin{tabular}[c]{@{}c@{}}optimizing routing\\ and storage \\ allocation\end{tabular}                       \\ \hline
		\begin{tabular}[c]{@{}c@{}}Jin et al.\\ ~\cite{jin19reinforcement}\end{tabular} & UWSN                                                                           & reinforcement                                                                  & QL                                                                      & \begin{tabular}[c]{@{}c@{}}information of\\ neighbors and\\ links\end{tabular}           & \begin{tabular}[c]{@{}c@{}}packet\\ forwarding\end{tabular}                   & \begin{tabular}[c]{@{}c@{}}constant cost, \\ congestion cost,\\ delay, and energy\\ consumption\end{tabular} \\ \hline
		\begin{tabular}[c]{@{}c@{}}Huang et al.\\ ~\cite{huang20resilient}\end{tabular} & \begin{tabular}[c]{@{}c@{}}wireless\\ sensor\\ network\end{tabular}             & supervised                                                                     & CNN                                                                     & \begin{tabular}[c]{@{}c@{}}adjacency \\ matrix\end{tabular}                              & \begin{tabular}[c]{@{}c@{}}link \\ reliability\end{tabular}                   & \begin{tabular}[c]{@{}c@{}}optimizing\\ routing\end{tabular}                                                 \\ \hline
		\begin{tabular}[c]{@{}c@{}}Zhang et al.\\ ~\cite{zhang20deep}\end{tabular}      & \begin{tabular}[c]{@{}c@{}}relay \\ network\end{tabular}                        & reinforcement                                                                  & QL                                                                      & \begin{tabular}[c]{@{}c@{}}states of\\ link, buffer,\\ and battery\end{tabular}          & link selection                                                                & \begin{tabular}[c]{@{}c@{}}maximizing\\ receiving data\end{tabular}                                          \\ \hline
		\begin{tabular}[c]{@{}c@{}}He et al.\\~\cite{he19route}\end{tabular}           & \begin{tabular}[c]{@{}c@{}}Cognitive\\ Radio\\ Network\end{tabular}             & reinforcement                                                                  & QL                                                                      & \begin{tabular}[c]{@{}c@{}}harvested energy,\\ battery,\\ destination nodes\end{tabular} & next hop                                                                      & optimizing energy efficiency                                                                                                \\ \hline
	\end{tabular}
\caption{Some Related Research Works Using AI to Optimize Transmissions for Green MTCs}
\label{ai-transmission}
\end{table*}

\subsubsection{Relay and D2D} Compared with routing, relay and D2D techniques provide more flexibility. AI can be adopted to decide whether to relay or not and help to select the optimal relay node according to the energy condition. Mastronarde et al. utilize the MDP to formulate the relay decision for each UE in the cellular networks~\cite{mastronarde16relay}. To maximize the long-term utility, the authors proposed a supervised learning-based model to help each UE to learn the optimal cooperation policy online. Specifically, the UE estimates three parameters, namely the outbound relay demand rate, inbound relay demand rate, and relay recruitment efficiency in an online manner. Then, the estimated values can be utilized to calculate the transition probability and utility functions. To address the problem of frequent recomputing, the authors first compute a collection of cooperation policies offline. Then, in the online phase, the estimated parameter values can be adopted to calculate energy cost, which finally helps to choose the optimal policy. 

He et al. study the relay selection problem in the air-to-ground VANETs (A2G VANETs) and adopt the QL to choose the relay node in order to balance the network performance and energy consumption~\cite{he20relay}. In this paper, the flying UAVs and the ground vehicles transmit  messages to each other by multi-hop relaying. Then, the relay selection affects the packet delivery ratio, latency, signal overhead, and energy consumption, which is further formulated to a multi-objective optimization problem. The authors construct the Q value table including the state and action indicating the network states and relay selection, respectively. Through attempting different relay selections, the Q values for different choices can be finally calculated. The extensive performance analysis illustrates the improvement in terms of packet delivery ratio, latency, hop counts, and signal overhead, which means increased energy efficiency. 

Wang et al. also utilize QL to optimize the power allocation and D2D relay selection for maximizing the  energy efficiency~\cite{wang20energy}. 
As the relay selection policy affects energy efficiency of all D2D pairs, the authors construct a finite MDP and adopt QL to choose which neighbor node is selected. In the QL model, the state space is defined with the four cases that whether energy efficiency of first-hop and second-hop D2D links is below or above the definite lower band. Each D2D pair acts as the agent to select a neighbor node in their region with the target of maximizing the reward defined according to their energy efficiency. Through the iteration process in the QL algorithm, the Q-value table of each D2D pair can be updated and the optimal relay with the maximum Q-value is chosen. The final simulation clearly illustrates the improvement of energy efficiency.

Hashima et al.~\cite{hashima20neighbor} utilize the stochastic MAB~\cite{auer02finite} to model the neighbor discovery and selection problem in mmWave D2D networks. And the considered MAB model aims to maximize the long-term reward which is defined as the average throughput of the devices subject to the residual energy-constraint of nearby devices. To solve this MAB problem, a family of upper confidence bound algorithm plus Thomson sampling is utilized by incorporating residual energy constraints. The final results illustrate the improved average energy efficiency and extended network lifetime. Authors of~\cite{abdelreheem19deep} also focus on the relay selection in D2D mmWave networks to increase the connection reliability. However, they utilize the DL model to predict the best relay device according to the distance between the device and BS or other devices, node mobility, signal strength, and residual energy. Specifically, the proposed relay selection algorithm consists of two phases. In the online phase, the random training values are generated with the best relay labels to train the considered DNN model. Then, the second phase is to utilize the trained DNN to predict the best relay. 

\subsection{Energy Harvesting and Sharing}
Similar to the cellular networks, MTC terminals can also be charged by the ambient energy in a wireless manner~\cite{mao20ai,zhao20deep}. To drive the MTC toward the green 6G era, two common energy harvesting techniques are expected to be widely applied: renewable energy harvesting and RF harvesting. The formal one considers renewable green energy sources such as solar, winding, tide, and so on to reduce the utilization of fossil fuel. The latter one is to efficiently harvest the dissipated energy which counts the majority in RF signals but cannot be used~\cite{zungeru12radio}. On the other hand, the dynamics of the harvesting power further complicate the network performance improvement or energy efficiency optimization, which is the reason for the application of AI techniques. In the following paragraphs, we introduce the related AI-based research considering the two EH techniques.

\subsubsection{Renewable Energy Harvesting}

Chu et al. utilize the RL technique to design the multiaccess control policy and predict the future battery state~\cite{chu19reinforcement}. In their research, the authors consider the uplink communication scenario where multiple energy harvesting-enabled UEs access the BS with the limited channel resource. The authors firstly assume the user battery and channel states are available for the BS, then utilize the DQN based LSTM to design the UE uplink access scheme. In this model, the system state includes the channel conditions and UE battery levels. The reward is defined as the discounted system sum rate of the long term. The consideration of multiple time slots drives the authors to adopt the LSTM model, which can make sequential decisions. The constructed LSTM model assists the BS to select the UEs at each time slot in order to maximize the system sum rate. In the second proposal, the authors utilize the RL based LSTM to predict the battery level. In this RL model, the considered state space includes the access scheduling history, the previous UE battery predictions, and the practical UE battery information. Since the purpose is to maximize the prediction accuracy, the reward is defined according to the long-term prediction loss. Finally, the authors combine the predictions of access control and battery information and design a two-layer LSTM DQN network. The first layer is to predict the battery level, which is adopted as part of the state space in the access control prediction. Extensive simulations illustrate the improvement of the system sum rate, further resulting in improved energy efficiency.

Similar to the considered scenario in~\cite{chu19reinforcement}, the same authors apply the DRL techniques to optimize the joint control of power and access~\cite{chu19power}. Generally, the proposal consists of two stages. In the first stage, the LSTM model is utilized to predict the battery states, which is similar to that in~\cite{chu19reinforcement}. In the control stage, the authors utilize the AC algorithm and DQN to decide the access and power scheme. The state space consists of the channel power gain, predicted UE battery level, history information of power control policy, and selected UE's true battery, while the action represents the transmit power which has a continuous value. The reward is defined according to the achieved transmission rate, thus the algorithm aims to improve the system throughput. The proposed LSTM model is verified a high accuracy rate to predict the battery state and the new approach enables the improved average sum rate compared with conventional algorithms as well as DQN-based models. 

From the above introduction, we can find that using the AI method to predict the harvesting-enabled battery state is an efficient method to adjust the network configurations for performance optimization. Authors of~\cite{mao19harvest,mao20ai} utilize the non-linear regression method to find the relationship between future harvesting power and the historical records. Then, with the estimated harvesting power, the IoT node can adjust the security configurations to provide qualified service as well as reduce the outage probability. In~\cite{mao20ai}, the authors also study the THz-enabled 6G IoT scenario and show the achieved network throughput improvement and extended working time.

\subsubsection{Radio Frequency Harvesting}
%~\cite{zungeru12radio}
Abuzainab et al. focus on the problem of adversarial learning in a communication network where the devices are served and powered by the Hybrid Access Point (HAP)~\cite{abuzainab17robust}. In the considered scenario, the HAP needs to estimate the transmission power of the devices and determine the suitable energy signal to reduce the packet drop rate of the devices. As the adversary may alter the HAP's estimate, the authors propose a robust unsupervised Bayesian learning method. In the proposed model, the HAP is assumed to have full CSI, which is utilized to calculate the transmission power according to the received signal power. In the nonparametric Bayesian learning model, the Dirichlet distribution is used to calculate the posterior distribution of the probability vector of the device transmission power. Then, the HAP can find the optimal transmission power to maximize the utility while not depleting the device's battery. Compared with the conventional Bayesian learning method, the proposed approach can achieve performance in terms of packet drop rate without jeopardizing energy consumption. The proposed learning scheme also exhibits improved energy efficiency compared with a fixed power transmission policy.

Kwan et al. study the RF harvesting from intended and unintended sources and propose machine learning-based wake-up scheduling policy for on-body sensors~\cite{kwan20coordinate}. To address the unpredictable nature and low amount of energy harvesting from the RF signals of unindented sources make it difficult to decide the wake-up time, the authors consider two machine learning techniques including linear regression and ANN to predict the wake-up time. In the linear regression-based forecaster, the authors consider the current capacitor charge level and average energy harvesting rate to address the dynamics caused by user mobility and changing channel conditions. The proposed ANN is to predict the next wake-up time considering the last successful wake-up time and energy level. The final simulation results illustrate the two models both achieve high accuracy rate.

Similar to~\cite{kwan20coordinate}, the authors of~\cite{yang19sample} also focus on the optimization of active time of IoT nodes which are powered by RF harvesting energy. In this paper, besides information collection and energy provision, the HAP is also responsible for setting the sampling time of the IoT devices. The challenge of this problem is that the HAP cannot have exact knowledge of the harvested energy for each IoT device due to the imprecise knowledge of CSI. To address this issue, the authors combine stochastic programming and RL techniques. Firstly, stochastic programming is used to maximize the minimum sampling time among all devices. To tackle the limitation of an unknown and dynamic probability distribution, the RL technique is adopted where the assumed agent decides the sampling and charging time according to the states corresponding to the device battery levels. The reward function is measured by the maximum-minimum active time of devices. Moreover, the authors model the large-state or continuous space using linear function approximation. The final results illustrate the RL approach can achieve as high as 93\% of the minimum sampling time computed by stochastic programming.

\subsection{Summary}
In this section, we analyze the AI-based research toward green MTCs. Compared with conventional methods, the advantage of AI is that it can address the uncertainty and alleviate the failure ratio during the access and transmission process~\cite{xu20priority,jiang19reinforcement,liu19reinforcement,wang20multi-hop,hu10machine}. For energy harvesting process, AI enables more knowledge about future available power and battery status, which enables necessary configurations towards improved energy efficiency~\cite{chu19reinforcement,chu19power}.

\section{Computing Oriented Communications}
\label{coc-section}
In the 6G era, the computation services are expected to play a more important role in people's work and life. With the great leap in transmission rate and communication capacity, an increasing number of applications will be offloaded to the cloud or edge server for the nearly real-time results instead of execution locally. Moreover, to store the contents on the cloud and edge servers can provide users with more efficient and flexible service. Additionally, the widespread application of AI techniques also drives the development of computing oriented communications to accelerate network management. In this section, we discuss the power consumption model and introduce the existing AI-based research aiming to improve energy efficiency and save energy consumption of the COC scenarios.
\subsection{Power Consumption Modeling}

The consuming power of the servers depends on the Central Processing Unit (CPU) or Graphic Processing Unit (GPU) utilization which usually keeps changing. Generally, energy consumption of a server is approximately linearly dependent on the CPU and GPU usage. If we assume $P_{idle}$ and $P_{max}$ to denote the consuming power of a server working at idle state and full state, respectively, the following equations can model energy consumption when the utilization rate is denoted as $u$~\cite{li18energy,wang13particle}:
\begin{align}
\label{coc-energy}
P(t)&=P_{idle}+(P_{max}-P_{idle})\times{u(t)}\\
E&=\int_t{P(t)dt}
\end{align}

Thus, for a cluster of servers, the total energy consumption can be calculated by summing energy cost of all servers. From Equations~\ref{coc-energy}, it can be found that to save energy consumption, we can reduce the utilization rate of each server. However, it has been investigated that the server in the idle state consumes approximately more than 60\% of the peak load electricity~\cite{fan07power,gupta11analysis}, which makes the problem more complicated. For a given workload, to utilize only one or sever servers at the full state and turn off the other servers may result in low energy consumption, but on the other hand contribute to the high delay. Therefore, how to allocate the computation resources to balance energy consumption and service quality is an important direction in the research~\cite{tian19reinforcement,zhang20dynamic,li20deep}. 

\subsection{Energy-Efficient Cloud and Edge Computing}
According to Equation~\ref{coc-energy}, to reduce the CPU/GPU usage can alleviate energy cost. In this part, we discuss the three common issues to alleviate the computation resource usage including offloading decision, resource allocation, and server placement. %The common strategies t
\subsubsection{Offloading Decision}
The existing networks usually consist of multiple types of computation platforms including the cloud, fog, and edge computation servers. Moreover, the computation tasks can be also executed locally if necessary. The heterogeneous computation platforms have variable latency as well as different energy consumption. Moreover, computation offloading choice also means different communication overhead. In this part, we introduce how AI is utilized to decide the computation offloading policy for green communications.

Wang et al.~\cite{wang19deep} combine the heuristic algorithm and DL to optimize the computation offloading policy to the fog or cloud servers. 
In this paper, the authors analyze energy consumption and latency to finish the computation task by fog servers and cloud server, and then formulate a Mixed Integer Non-Linear Programming (MINLP) problem aiming at minimizing the total energy cost under the latency constraint. To solve the NP-hard problem, the authors first utilize the simulated annealing algorithm~\cite{simulated-annealing-wiki} to find some optimal solutions, which is further utilized to train the constructed CNN model. Moreover, the training process is periodically conducted to update the parameters of the considered CNN models, while the greedy algorithm is utilized as compensation if the result of CNN models is not reasonable. 

\begin{figure}[!t]
	\includegraphics[scale=0.3]{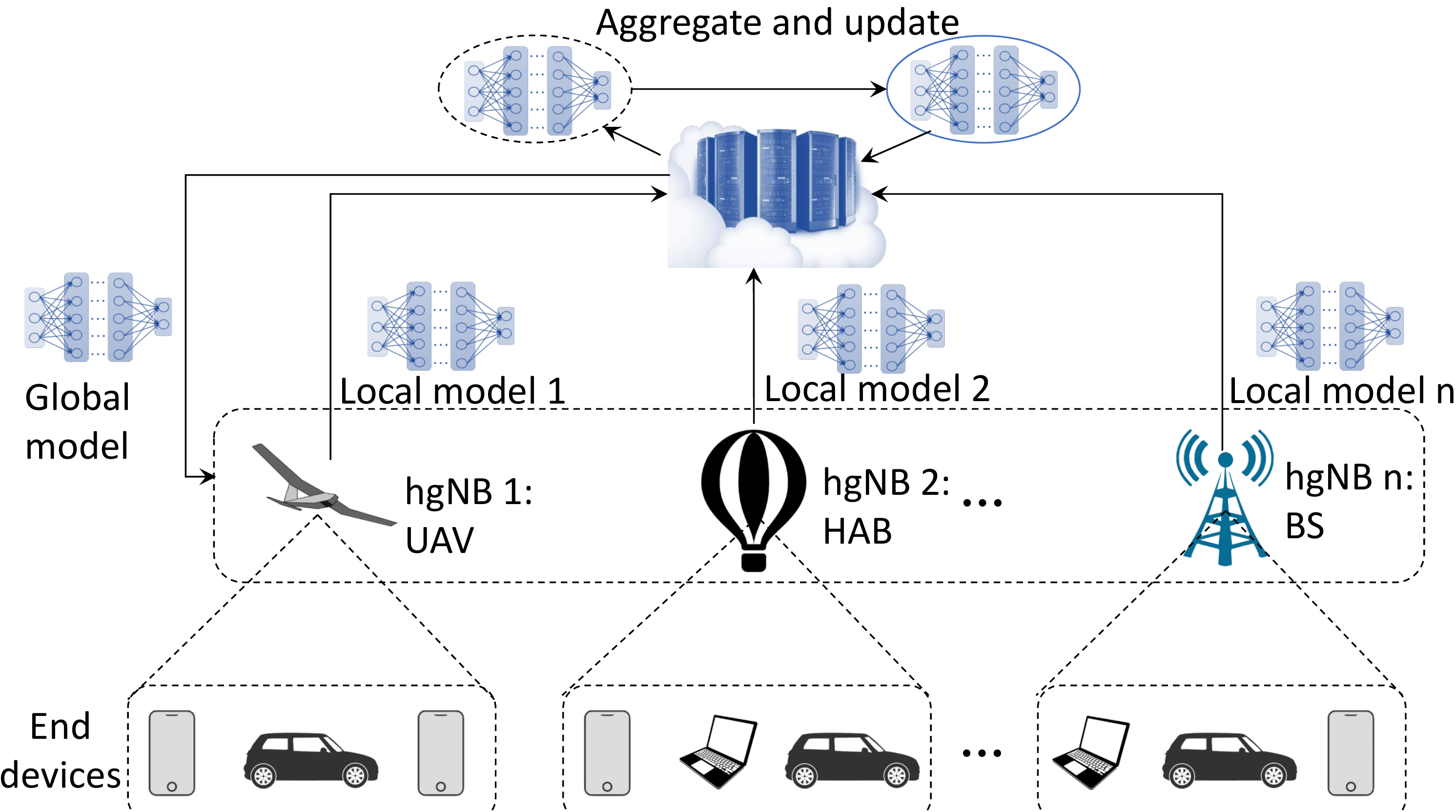}
	\caption{The federated learning structure for heterogeneous network architecture.}
	\label{fl-model}
\end{figure}

To alleviate the computing overhead for meeting the latency requirement, Gong et al.~\cite{gong20deep} consider the high-rate RF communication and low-power backscatter communications to realize active offloading and passive offloading, respectively. Since the local computing, active offloading, and passive offloading have different computation latency as well as various energy consumption, DRL is adopted to optimize the suitable computation and transmission policy. The assumed agent chooses from three actions: local computing, active offloading, and passive offloading, given the channel conditions, energy status, and workload in each time slot. The final results illustrate that the DRL-based method can reduce the outage probability by intelligently scheduling the offloading policy. Moreover, this paper illustrates many perspective directions of DRL-based backscatter-aided data offloading in Mobile Edge Computing (MEC) scenarios. 

Yan~\cite{yan20offloading} consider the single user with multiple independent tasks meaning that the results of some computing tasks may be utilized as the input of some others. The authors first adopt the task call graph~\cite{kwok96dynamic} to model the inter-dependency among different computation components. To reduce energy consumption of mobile devices and the computation latency, the authors define the reward function with weighted energy consumption and latency, then consider two problems: how to offload the computation tasks and how to allocate the CPU cycles for different tasks. Since the first problem is a combinatorial binary problem while the second one is convex, the authors adopt the DRL technique to decide the offloading policy. The AC learning structure~\cite{actor-sergey} is utilized where a DNN in the actor network is adopted to learn the relationship between the input states (wireless channels and edge CPU frequency) and the offloading policy, while the critic network is to evaluate the energy and latency performance of different offloading strategies. Different from the conventional critic network which utilizes the DNN to evaluate the offloading decisions, the authors define the one-climb policy where the tasks in one path of the task graph can only migrate once from the mobile device to edge servers, which can reduce the number of performance evaluations, resulting in reduced complexity and accelerated computations of the DRL method.

Ren et al.~\cite{ren19federate} unite the federated learning and DRL to optimize the partial offloading policy for the energy harvesting-enabled IoT nodes. Different from the centralized learning technique, the adopted federated learning enables every IoT node to avoid the sensing data uploading to the edge node, which can protect data privacy and alleviate the transmission overhead. Specifically, the edge node only acquires the parameters of the trained DRL agent, while a random set of IoT devices are selected to download the parameters from the edge node, train the DRL agent with newly collected data, and finally upload the updated parameters to the edge node. In this paper, the authors also compare with centralized DRL in terms of training performance and network performance. Results show that the training of FL-based DRL can finally approach that of centralized DRL, even though it fluctuates more seriously. And under varying computation task generation probability, the federated learning-enabled DRL can improve the overall network performance, especially in terms of queuing delay and task drop rate. Similar research is also conducted in~\cite{shen20computation}. The authors demonstrate that the federated learning-based DRL models can be applied to various environments with reduced transmission consumption and enhanced privacy protection.

Different from the above works focusing on static scenarios, the authors of~\cite{wang20imitation} adopt multiple ML techniques to optimize the cooperative Vehicular Edge Computing (VEC) and cloud computing in dynamic vehicular networks. As the uncertain vehicle mobility results in the dynamic network structure and unstable connections, which leads to low efficiency for conventional heuristic searching strategies, ML is adopted~\cite{zhang19new} to cluster the vehicles into groups according to their connection time, where each group consists of a Road Side Unit (RSU), multiple service demanding vehicles, and service providing vehicles. And the RSU decide whether offloading the tasks to the cloud servers or conduct them locally. To schedule the computation tasks for a balance of energy consumption and latency, an imitation learning-based algorithm is proposed, which can alleviate the extreme complexity of conventional branch-and-bound algorithm. Specifically, an expert is trained with a few samples to obtain the optimal scheduling policy in an offline manner. Then, the agent is trained to follow the expert's demonstration online. Results illustrate that imitation learning can significantly accelerate the execution of the branch-and-bound process. 

\subsubsection{Computation Resource Allocation}
The computation platform usually needs to execute multiple tasks. How to allocate the computation resource, especially the CPU/GPU cycles is an attractive topic~\cite{wang20imitation,pradhan20computation,wang2020federated,ma20joint}. On the other hand, energy consumption is also an important metric that needs to be considered. How to balance energy consumption and computation performance can be addressed by AI techniques~\cite{wang20imitation,wang20computation,cheng19space}. The following paragraphs will focus on several research works.

Similar to~\cite{wang20imitation}, the authors of~\cite{xu17online} also consider AI techniques to balance energy consumption and latency for the scenarios utilizing the capacity-limited edge servers and cloud server. However, edge servers are driven by hybrid power including solar, wind, and diesel generator, while computation-efficient cloud servers are grid-tied. The authors model the joint workload offloading and edge server provision as an MDP and utilize the RL technique to solve it. The authors define the total system cost with the delay, diesel generator cost, and battery consumption, while the policy denotes the computing power demand in each time slot. To find the optimal policy, a novel post-decision state-based online learning algorithm is proposed to exploit the state transitions of the considered energy harvested-enabled MEC system. Compared with the standard QL method, the proposed approach converges much faster. And extensive simulations confirm that the MEC system performance can be significantly improved.

Pradhan et al.~\cite{pradhan20computation} study the computation offloading of IoT devices in the massive MIMO Cloud-RAN (C-RAN) deployed in an indoor environment. In this paper, the purpose of optimizing the computation offloading is to minimize the total transmit power of IoT devices. In the considered scenario, the transmission latency of the uplink signals is concerned with the transmit power and the CPU cycle allocation.  Therefore, to minimize the total transmit power of IoT devices under the latency threshold, we need to consider not only the signal processing factor, but also the computation resource allocation, which is a non-convex problem due to the coupling relationship among these factors and their value constraints. To solve this problem, the authors consider the supervised learning method and adopt the DNN model to decide the transmit power, CPU cycle assignment vector, and the number of quantized bit. The authors also propose an Alternating Optimization (AO) based mathematical model to obtain some near-optimal solutions to train the DNN model offline. Simulation results illustrate the fast convergence of the DNN training process. More importantly, to tackle the same problem in dynamic IoT networks, the authors utilize the transfer learning~\cite{pan10survey} technique, which means that part of the trained DNN's parameters are utilized in the newly-formed DNN for the changed scenario. Then, the DNN can be updated through training with limited samples, avoiding the complex training from scratch, which reduces the execution time by the order of two magnitudes. The final performance analysis show that the transfer learning-based DNN can provide a close approximation of the optimal resource allocation.

Wang et al.~\cite{wang2020federated} study the cellular networks where MEC-enabled High-Altitude Balloons (HABs) conduct the users' computation tasks with limited capacity and energy. Since the data sizes of the computation tasks vary, the user association policy should be optimized to meet the requirement as well as minimize energy consumption. To alleviate the limitations of traditional Lagrangian dual decomposition~\cite{zhang17energy} and game theory~\cite{moon16energy} in dynamic scenarios, the authors utilize the SVM-based federated learning algorithm to map the relationship from users' association and historical requested task size to the future association. 
Specifically, similar to the process in Fig.~\ref{fl-model}, the HAB first train an SVM model with the locally obtained data to construct the relationship between user association and computation task size. Then, the HABs share their trained SVM model, which enables further integration and local improvement. Thus, each HAB can build an SVM model to quantify the relationship between all user association and historical computation task information. The simulation results illustrate energy consumption can be reduced with a better prediction of optimal user association.

Ma et al.~\cite{ma20joint} utilize the PSO algorithm to jointly optimize the selection of access networks and edge cloud to minimize the latency and total energy consumption. In the considered scenario, each user can be served by multiple edge cloud-enabled access networks. Since the latency and energy consumption are both caused by task offloading and execution, the formulated problem to minimize the two metrics is NP-hard. In the adopted PSO model, the fitness function is defined as the sum of weighted latency and energy consumption. Note that the values of latency and consumed energy are processed to between 0 and 1 to avoid the dimensional influence. And the final performance analysis illustrates the significant improvement in terms of latency and energy consumption.

\subsubsection{Edge Server and Virtual Machine Placement}
The placement optimization including the edge servers and Virtual Machines (VMs) affect the resource utilization of the whole network. Since power consumption at the idle state constitutes the major part of total energy waste~\cite{liu18energy}, to minimize the active servers as well as meet the service requirements can improve energy efficiency. And AI techniques including the heuristic algorithms and machine learning methods have been studied to optimize the deployment of edge servers and VMs.

Li and Wang~\cite{li18energy} study the edge server placement and devise a PSO-based approach to minimize energy consumption. In this paper, the authors consider that multiple edge servers are located at different base stations. And the delay for the base stations to access the edge servers should be not above a threshold. In this paper, the minimization of energy consumption depends on the locations and assignments of the edge servers. To solve this discrete problem, the authors also redefine the parameters and operators of the PSO method. To evaluate the performance, a real dataset from Shanghai Telecom is utilized in the experiment, with which the PSO-based approach shows an improvement of more than 10\% energy saving.

Liu et al.~\cite{liu18energy} study the VM placement in cloud servers and adopt the ACO algorithm to minimize the number of active servers and balance the resource utilization, resulting in improved energy efficiency. In their approach, the bipartite graph is constructed to describe the VM placement problem. And the pheromone is distributed not only between the VMs and servers, but also among the VMs assigned to the same server. 
And the assumed artificial ants conduct the VM assignment based on global search information. To speed the convergence and improve the solution, a local search including the ordering exchange and migration operations is conducted. The improved ACO algorithm is efficient for large-scale problems. And the experimental results show that the number of active servers can be minimized with balanced usage of resources including the CPU and memory, which results in improved energy efficiency.

Shen et al.~\cite{shen19energy} focus on the cloudlet placement to improve energy efficiency in the mobile scenario and K-means clustering~\cite{visalakshi09kmeans} method is adopted to search the location center. In this paper, energy consumption is assumed to be directly related to the number of deployed cloudlets. Thus, to minimize the number of deployed cloudlets can optimize energy efficiency. To tackle this problem, the authors firstly utilize the K-means clustering method to find the central locations of the mobile devices. The following steps are to delete some locations that do not meet the density requirements and generate the moving trajectory of the cloudlets. Performance analysis illustrates the increased number of covered devices of each cloudlet, which results in reduced energy consumption.

Zhang et al.~\cite{zhang19genetic} study the container placement to optimize energy consumption of virtual machines and propose an improved GA. In this paper, the container is utilized to compute some applications and energy consumption is assumed to be nonlinearly related to resource utilization. Since the container placement is regarded as a combinatorial optimization problem, the heuristic algorithms, such as GA~\cite{gong18set}, are well suited. However, the conventional GA sometimes incorrectly eliminates new individuals in the mutation operation when resource utilization is high, which causes performance degradation. To solve this problem, the authors propose two kinds of exchange mutation operations and define a control parameter with the number of search iterations. And the method can help the search iteration to jump out of the local optimum. The final simulations illustrate the significantly improved power saving performance in small, medium, and large scales of scenarios with uniform and non-uniform VM distributions.

Wang et al.~\cite{wang13particle} study virtual machine placement in heterogeneous virtualized data centers and utilize the PSO method~\cite{holland84genetic} to minimize energy consumption. In this paper, the authors first establish energy consumption model of a heterogeneous virtualized data center. Since traditional PSO method can be only utilized for continuous optimization problems, the authors redefine the particle position and velocity with two $n$-bit vectors, and then redefine the subtraction, addition, and multiplication operators to fit the energy-aware virtual machine placement optimization, which is a discrete problem. Then, the authors consider the energy-aware local fitness and devise a two-dimensional encoding scheme to accelerate the convergence and reduce the search time. Results illustrate that the proposed method outperforms the other approaches and can lessen 13\%-23\% energy consumption. A similar research work based on PSO is given in~\cite{ibrahim20power}. The authors utilize the decimal coding method to apply PSO in a discrete problem. And energy consumption is minimized considering the service requirement constraints. The authors also analyze the complexity of the proposal which is related to the numbers of migrated virtual machines, particles, and iterations. 

\subsection{Green Content Caching and Delivery}
Besides offloading the contents to the edge/cloud servers, to store the contents is also an important service for future CDN. Energy consumption of this part mainly comes from the caching and delivery process. In the following paragraphs, we discuss the related research on how AI is adopted to improve energy efficiency of content caching and delivery.

\subsubsection{Caching Policy Design} For future multi-tier or hierarchical networks, the contents are usually cached in different parts to improve storage efficiency. The content caching policy needs to be optimized due to the variable storage size of heterogeneous devices and different energy consumption for content retrievers. Li et al.~\cite{li19reinforcement} utilize the DRL to optimize the content caching policy for multi-tier cache-enabled UDNs. The authors analyze the different energy consumption of content retriever from the Small Access Points (SAPs), MBS, and core networks, then construct the energy-efficient model. To optimize energy efficiency, the standard DRL method using the regular multi-tier DNN is adopted, where energy efficiency and different content combinations as the reward and state, respectively. To accelerate the convergence of the proposed intelligent content caching method, the authors utilized the latest finds including the prioritized experience replay~\cite{schaul16prioritiz}, dueling architecture, and deep RNN. Extensive simulations illustrate that the proposed intelligent content caching algorithms can significantly improve energy efficiency for both the stationary and dynamic popularity distributions. \cite{somuyiwa18reinforcement} analyzes impacts of the channel conditions on content caching. And the RL-based content caching is proposed to alleviate energy consumption.

Shi et al.~\cite{shi20novel} adopt the DQN model to optimize the content caching in three layered vehicular networks, where an airship distributes the contents to UAVs for satisfying the terrestrial services. In the considered scenario, the airship needs to schedule the UAV caching the required contents to provide the service if the requested content is not in local UAV, which means more energy consumption. To minimize energy consumption, the DQN model is proposed and the defined reward considers the probabilities of local UAV requests and other UAV scheduling. To improve training performance, the experience replay mechanism is considered. And the proposed DQN model is verified to overcome the large number of states and  in the training process.

Tang et al.~\cite{tang20energy} consider the scenario where the users can retrieve the contents locally, or from the neighbor devices, SBS, and MBS, with increasing energy consumption. On the other hand, the user's device, SBS, and MBS have increasing caching capacity. Specifically, the QL algorithm is applied to every user to select the cached contents with the goal of minimizing the cost which is inversely proportional to the popularity of cached files. For the caching policy of each SBS, the DQN is adopted to select the contents in order to minimize the total energy consumption. In the proposal, the cost function is similar to the reward in DRL, while the optimization goal becomes to minimize the value of cost. For this proposal, the complexity of QL is relatively low since every user's device has very limited capacity, which means the state space is small. On the other hand, the DQN has a relatively high complexity since the number of cache combinations is large, leading to a huge state space.

The content caching policy design deeply depends on the users' preferences, thus, the centralized control-based optimization methods may cause concern for privacy. For the data-driven AI algorithms including ML and DL techniques, the training and running process which requires the users' data poses great challenges. To address this problem, federated learning has been widely studied to keep the data IN the local area to protect privacy~\cite{fadlullah20heterogeneous,yu18federate,cui20blockchain}. In~\cite{fadlullah20heterogeneous}, the UE conducts the calculations of the shallow layers to generate some general features of the content requests. Similar to the process in Fig.~\ref{fl-model}, the heterogeneous BSs including the flying UAVs aggregate the parameters of the shallow layers to conduct the further training and running process to decide the content caching policy. Different from the cooperative training of the deep learning models, Yu et al.~\cite{yu18federate} consider that each user downloads the Stacked Autoencoder from the server and trains it with the local dataset generated from the personal usage. Then, the updated parameters and extracted features are uploaded to the server, where the hybrid filtering technique is adopted to decide the content caching policy. To further ensure data security, blockchain techniques can be adopted in the data transmission process~\cite{cui20blockchain}. However, these research works aim to improve the caching performance, instead of the minimization of energy consumption. 

\subsubsection{Delivery} Besides content caching, how to deliver the contents is also an important factor to affect energy consumption. In this part, we discuss the related AI-based research on content delivery optimization.

Lei et al.~\cite{lei17deep} study the content caching and delivery in cellular networks, and a supervised DNN based approach is adopted to optimize the user clustering to minimize the transmit power of the BSs. In each cell, the content delivery should satisfy the stringent delay requirement, thus the user scheduling algorithm should have low computation time to enable real-time operations. To realize this goal, the DNN is trained to map from the users' channel coefficients and requested data amount to the clustering scheduling policy. The authors utilize a variable size of dataset generated with conventional iterative algorithms to train the proposed DNN. And the performance shows that the large sized dataset can result in 90\% approximation to the optimum with limited time consumption. 

Al-Hilo et al.~\cite{al-hilo20uav} utilize the DRL technique to optimize the trajectory of UAV in order to improve content delivery for the UAV-assisted intelligent transportation system. In this paper, the moving vehicles are assumed to cache part of the contents due to the limited capacity and need to retrieve the other contents from the BS which is time-consuming and unstable. To improve the content delivery performance, the cache-enabled UAVs are assumed to hover over the vehicles to meet some content requests. As the trajectory control affects the performance of content delivery as well as the power consumption of UAVs, the Proximal Policy Optimization algorithm is adopted to decide the flying velocity according to the network states including the current position, vehicle information, and cached contents. The final results also show the improvement of energy efficiency. 

The above works focus on content delivery in the access networks, while the data forwarding in the core networks is also an important factor to affect energy consumption. Li~\cite{li15energy} utilize the ACO algorithm~\cite{dorigo07ant} to optimize the data forwarding scheme to reduce content retrieve hops, which results in less energy consumed by the routers and links. In this paper, the CDNs are first divided into multiple domains. And the data packets and the hello message packets are assumed to be two types of ants. For each path, the pheromone is defined and calculated as the normalized sum of path load, delay, and bandwidth. Then, through the generated interest ants in the initial state, the node can construct the paths and update the corresponding pheromone values. Then, during the data packet transmission stage, the pheromone is further updated according to the real-time performance. 

\subsubsection{Joint Optimization} Since the caching and delivery policies both affect energy consumption, joint optimization is another direction toward green communications. Li et al.~\cite{li20green} adopt the DRL method to minimize the latency and energy cost of content caching and delivery in RAN. In this paper, the authors define the reward function considering the latency and energy cost of the content caching and delivery between the users and SBS, MBS, and cloud servers. Then, the AC model and DDPG algorithm~\cite{guha20deep} are adopted, where two identical DNNs are utilized to generate the deterministic action and evaluate the chosen strategy. Here, the action is defined with the content file placement, SBS-user association, and subchannel assignment. The simulation results illustrate the improved rewards, which means the performance improvement in terms of transmission delay and energy consumption.

Similarly, Li et al.~\cite{li20deep} also utilize the DL technique to jointly optimize the content delivery latency and system energy consumption. However, as the cache-enabled D2D networks are adopted to alleviate the overhead of requesting the contents from the cellular BS In this paper, the device mobility, content popularity, and link establishment decisions need to be considered. To address the complexity caused by the dynamics including changing channel conditions and variable content popularity, the authors consider a three-step proposal, all of which utilize the DL models. First, the RNN models including the conceptor-based Echo State Networks (ESN)~\cite{l12practical} or LSTM is utilized to predict user mobility according to the limited previous records. Then, the predicted D2D user location information, together with other attributes including gender, age, occupation, time, and so on, are utilized as the input of ESN or LSTM to predict the probability of each user to request every content at the next time slot. Then, the content request distribution can be utilized to assist the content placement. For example, the content will be assigned to the user if the request probability is above 70\%. In the third step, the joint value and policy-based AC algorithm~\cite{actor-sergey} is utilized for each user to choose a neighbor to establish the communication link for content delivery according to the observed environment which is defined as the transmit power, channel gain, and distance. In this algorithm, the reward function is denoted by the sum of weighted content delivery delay and power consumption. The simulation results illustrate that with different weight combinations of delay and power consumption, variable power saving performance can be obtained, which means that the proposed strategy is reasonable and flexible. Similar research is given in~\cite{chen17cache}, which also utilizes the ESN model~\cite{l12practical} to predict the user mobility and content request distribution. Since the requested content is dependent on the users, the authors consider the context of users including the gender, occupation, age, and device type to predict the probability of content requests. To make the results practical, the authors collect historical content transmission and user mobility records to train the considered models.
 
\subsection{Summary}
According to the introduced research, we can find AI techniques can significantly improve energy efficiency of the content caching process. In the content placement step, AI techniques are important and efficient to predict the content popularity and users' information including the preference and location, which can result in improved local Cache Hit Ratio (CHR) and reduce the content retriever from cloud servers. For the content delivery part, the optimization is to improve the resource allocation, transmission scheduling, routing, and other communication functions to save energy. Different from the energy-efficient proposals in cellular networks as we mentioned above, the strategies in content delivery networks should consider the content placement, latency requirements, and even the caching capacity.

\section{Open Research Issues}
\label{open-research-section}
Even though there are a huge number of research works on AI-based green communication services, we still need to pay more attention to transform our endeavors into practical applications in the 6G era. Moreover, the utilization of AI techniques in current networks is still confronted with many challenges in terms of computation complexity, hardware compatibility, data security, and so on. The following paragraphs give some promising directions, which we believe will give some ideas to the researchers.

\subsection{Green BS Management for 6G HetNet}
As we mentioned in Sec.~\ref{cnc-section}, the BSs take the majority of total energy consumption. In the 6G era, the number of BSs is meant to be multiple times that of 5G. And these BSs are constructed in a hierarchical manner and have various sizes of coverage. Moreover, as the UAVs and HABs will also act as the BSs~\cite{cheng19space,al-hilo20uav,wang2020federated,fadlullah20heterogeneous}, the heterogeneous hardware architectures and the mobility further complicate the green management. The following paragraphs introduce the potential AI-based research considering the potential three functions of 6G BSs.

As the end terminals can be served by different BSs including the MBSs, SBSs, and Tiny Base Stations (TBSs) in the multi-tier 6G HetNet, the user association policy should be optimized in order to turn off the redundant BSs for energy saving. Moreover, the BSs are usually deployed with multiple frequency bands, the resource allocation including the channels and power are critically for the network energy efficiency. However, the mobility of end devices, and UAV or satellite-enabled BSs results in the changing traffic demand and dynamic channel conditions, while the resource heterogeneity further complicates these problems. To address these issues, AI techniques can provide efficient assistance. For example, AI models can be adopted to predict the traffic demands, mobility patterns, and channel conditions, which enables the network reconfigurations in advance. 

Besides offering communication services, future BSs will act multiple roles, such as the computation/storage providers and energy source. As some BSs have a certain amount of computation and storage resources, the computation offloading and content caching policies can be optimized by AI models. For example, the computation offloading or content caching are usually models as a non-convex problem, which is further solved by the RL or DRL techniques. As we mentioned in Sec.~\ref{intro}, compared with the traditional method which divides the non-convex problem into two sub-problems and solves them one by one, the RL or DRL can find the global optimal solution and avoid the complex iteration process during the algorithm execution period. 

\subsection{Energy-Efficient Space-Air-Ground Integrated Networks}
SAGIN has been regarded as one of the key technologies for 6G~\cite{khaled19roadmap,liu18space}. SAGIN can provide seamless coverage and flexible information transmissions, especially for massive MTCs. Since the satellites, HABs, and many UAVs are driven by renewable energy, energy-efficient network orchestration is critically important for SAGIN. However, the diversified transmission environments, heterogeneous hardware platforms, and dynamic energy resources pose great challenges. To address the complexity and uncertainty, AI can provide many efficient models. For example, using the RL technique to optimize the resource allocation policy including the transmitting power~\cite{tsuchida20efficient} and channels~\cite{zhao20deep} has been evaluated to improve the network energy efficiency. Moreover, the CSI dynamics and network mobility make energy-efficient packet transmissions more difficult. As AI has been demonstrated that it can efficiently map the complex relationship between existing network traces and future transmission policy for terrestrial networks~\cite{zhao20deep,zhou18energy}, we believe the research can be extended to the SAGIN scenario.

Even though AI has been studied to optimize the SAGIN performance~\cite{kato19optimize,zhou20deep}, current research mainly focuses on the single layer, such as the LEOs and UAVs. From the systematic perspective, the network management toward green communications should consider every part of SAGIN. For example, the UAV deployment and trajectory should be optimized considering the beam control of satellites to realize energy-efficient coverage~\cite{liu18energy,cao19deep,yuan20energy}. As AI has been illustrated to be competent to handle the complex multiple-variable-related problems~\cite{yang20learn,yang20actor}, using AI techniques to analyze performance from the perspective of whole SAGIN system will be a promising direction. However, the difficulty is how to characterize the concerned factors into the AI model~\cite{kato17deep,mao18novel}. And, the execution of the AI model is another challenge due to the extreme computation overhead. Moreover, AI is also important to optimize RF energy harvesting in cellular networks, which will be discussed in Sec.~\ref{future-eh}.

\subsection{AI-based Energy-Efficient Transmissions}
Packet transmission is energy-consuming as it costs energy of transmitters, forwarders, and receivers. Besides power control and resource allocation methods to reduce energy consumption, many other choices have been provided including the routing policy design, relay, backscatter communication, and IRS-aided transmissions. There is no doubt that multiple communication manners will be provided for the end devices to transmit the packets successfully. For instance, the mobile users can choose the cellular network to send the email, which can be also finished by the IEEE 802.11-based WiFi or through D2D in a multi-hop manner. How to cooperatively utilize and schedule the different communication methods and resource in a multi-agent multi-task environment will heavily affect the system energy consumption and network performance. Most AI-based research focuses on the single communication scenario, while very limited works study the hybrid scenario~\cite{zhang20energy3,yang20learn}. In the future, we can pay more attention on AI to improve energy-efficient transmission in the scenario where multiple communication manners are available.

\subsection{AI-Enhanced Energy Harvesting and Sharing}
\label{future-eh}
Energy harvesting has been widely recognized as an important part for green communications. To drive the development of green communications, various energy harvesting techniques will be utilized, which can be grouped into different groups according to whether it is controllable and predictable~\cite{kansal07power}. AI techniques can be adopted in the scenarios using the uncontrollable but predictable energy group and partially controllable energy group, where the formal consists of the solar, winding, tide, and other renewable sources, while the latter includes RF energy. For the uncontrollable but predictable energy harvesting techniques, some AI models can be utilized to map the relationship between the future harvesting power and related factors~\cite{gambin20share,jahangir20deep}. And the predicted results can be adopted to reconfigure the network in advance. Another method is to directly utilize AI models to map from the harvesting-related factors to network management policy. These methods enable network operators to gain more knowledge of energy harvesting and improve the utilization efficiency. For the partially controllable RF energy harvesting technique, AI can be used to optimize the BS power control and transmission scheduling~\cite{liu19power,gao17qlearning}. For the UAV-enabled BSs, AI can be adopted to optimize the trajectory to reduce energy consumption and improve the harvesting efficiency~\cite{hoseini20trajectory,nie20energy}. Current research mainly focuses on the maximization of minimum harvesting energy due to the disordered transmission and unplanned power control~\cite{kwan20coordinate,yang19sample}, AI can enable the RF harvesting process to be energy-aware, which can greatly reduce the wasted energy, especially for the signals from omnidirectional antennas.

The RF harvesting technique also enables energy sharing among devices, which can be considered to avoid the outage of some network parts as well as reduce energy waste when batteries of some devices are nearly full and cannot save incoming energy anymore~\cite{lin15distribute}. The Simultaneous Wireless Information and Power Transmission (SWIPT) technique has been widely studied, especially in MTC scenarios~\cite{perera18simulateneous}. Even though it may cause some performance loss to harvest energy from part of the received signals, AI can be utilized to decide the ratio between RF harvesting and information transmission to reach a balance~\cite{liang20optimal}. Currently, ambient backscattering is a promising technique especially for the low power machines, AI can be considered to optimize energy harvesting and information forwarding process~\cite{zou20optimize,gong20optimize,gong20deep}.

\subsection{Security for AI-enabled Networks}
The adversaries and unauthenticated users threaten the information privacy as well as cause the transmission failures, leading to the deteriorated energy efficiency. To protect the normal information transmission from the attacks, AI can be considered as it has been verified to detect the network threats~\cite{cui18detection}. Moreover, using AI to control the transmit power and allocate the resource is also efficient to address the network jammers~\cite{han20dynamic}. For the future AI-driven 6G, a new type of network threatens may be the malicious data generated by the adversaries, which misleads AI models to reach a wrong decision. Besides the decreased throughput or increased latency, the potential results may be the widespread outage of end terminals or extremely low harvesting efficiency. How to develop robust AI models to ensure green communications will be important topics.   

Most AI techniques including the DL and ML rely on data in the training and running phases. Since the data may be concerned with personal privacy or business information, to develop and execute AI algorithms should consider the data security issues. More importantly, the standards and regulations should be built to guide the collection and usage of data~\cite{kato20ten}. 
\subsection{Lightweight AI Model and Hardware Design}
To develop AI-based green communications, energy consumption of AI algorithms should be analyzed. However, most of the current research just focuses on the network performance improvement compared with conventional algorithms and neglects the consumed energy for the training and running of AI models~\cite{chen20edge,nishio19client}. This may cause the high complexity of the proposed AI models, which may be more energy-aggressive than traditional methods. Thus, how to minimize the required training data and how to decrease the algorithm complexity is important for the development of AI-based green communications. As the reduced complexity may sacrifice the accuracy rate in some cases, the balance energy efficiency and network performance is still critical for AI algorithms. Furthermore, the amount of consumed energy for AI algorithms also depends on the hardware. To design the hardware for computation acceleration of AI algorithms with low cost should also be paid more attention~\cite{yamauchi19implement}. Currently, very limited research has analyzed how to conduct AI algorithms with low energy consumption~\cite{yang20energy,eshratifar19efficient}. And the results inspire us to pay more attention to how to execute the proposed AI algorithms in an energy-efficient manner.

\balance

\section{Conclusion}
\label{conclusion}
AI has aroused widespread attention from nearly every field to improve the quality, accelerate production, customize the provided services, and so on. To utilize AI technologies in 6G has been widely acknowledged as a paradigm. And the AI-based green communications will be an important direction due to the exponentially increasing energy consumption from the growing infrastructure and end devices. To reduce energy cost and improve energy efficiency, too many variables and a high dimension of solution space need to be considered and analyzed. Conventional heuristic algorithms and convex optimizations require the simplification of considered problems, which may need a great number of iterations or not reach a satisfying energy efficiency level. On the other hand, AI techniques have been verified their overwhelming advantages and power in handling complex problems. In this research, we survey the AI-related research on network management and configurations toward energy efficiency optimization. Another direction for green communications is to utilize energy harvesting techniques which adopt renewable energy or ambient energy to reduce the usage of fossil resource. AI techniques can be adopted to address the uncertainty and dynamics in energy harvesting process. Moreover, this paper considers three common scenarios in 6G: CNC, MTC, and COC, and analyze how AI can improve the configurations of 6G elements including massive MIMO, NOMA, SAGIN, and THz. We believe this paper can provide some guidance and encourage future works focusing on AI-based 6G green communications.

Furthermore, we analyze the strengths and weakness of different AI models, including the traditional heuristic algorithms and the state-of-the-art ML/DL methods. We illustrate how they can cooperatively work to reduce energy consumption and improve energy efficiency from a systematic perspective. Additionally, we discuss the necessity to consider energy consumption of AI models and indicate some open issues including data privacy, computation complexity, hardware design, and network deployment, which the future researchers need to embrace.

%\balance

\Urlmuskip=0mu plus 1mu\relax

\bibliography{myRef}{}

\bibliographystyle{ieeetr}

\end{document}